\documentclass[iop]{emulateapj}

\usepackage{graphicx}
\usepackage[singlelinecheck=on,caption=false]{subfig} 
\def\wcen{$\omega$~Cen}
\def\lx{L$_x$}
\def\fx{${f}_X$}
\def\fr{${f}_R$}
\def\fxfr{${f}_X$/${f}_R$}
\def\fratio{${f}_X$/${f}_R$} 
\def\fxfopt{${f}_X$/${f}_{opt}$}

\def\a{$\&$}
\def\x{$\times$}
\def\about{$\sim$}
\def\simlt{\buildrel{<}\over \sim}
\def\simgt{\buildrel{>}\over \sim}
\def\simgreat{\buildrel{>}\over \sim}
\def\simlt{$\la$}
\def\simgt{$\ga$}
\def\asec{$''$}
\def\amin{$'$}
\def\secspt{$\buildrel{\prime\prime}\over .$}
\def\minspt{$\buildrel{\prime}\over .$}

\def\msun{${\cal M}_{\odot}$}
\def\lsun{$L_{\odot}$}
\def\sun{$_{\odot}$}
\def\yr-1{yr$^{-1}$}

\def\bwfpc2{$B_{439}$}
\def\b{$B_{435}$}

\def\r{$R_{625}$}

\def\br{$B_{435}-R_{625}$}

\def\hrwfpc2{$H\alpha-R_{675}$}
\def\hr{$H\alpha-R_{625}$}

\def\Mr{$M_{625}$}
\def\ha{H$\alpha$}

\def\HST{$\it HST$}
\def\Chandra{$\it Chandra$}

\def\ergs{erg s$^{-1}$}
\def\ergscmsq{erg s$^{-1}$ cm$^{-2}$}
\def\NH{N$_{\rm H}$}

\received{September 14, 2012}
\accepted{December 8, 2012}

\begin{document}

\title{HST/ACS Imaging of Omega Centauri: Optical
Counterparts of Chandra X-Ray Sources}

\author{Adrienne M. Cool,\altaffilmark{1}
Daryl Haggard,\altaffilmark{2,3}
Tersi Arias,\altaffilmark{1,4}
Michelle Brochmann,\altaffilmark{1,5}
Jason Dorfman,\altaffilmark{1,6}
April Gafford,\altaffilmark{1,7} 
Vivian White,\altaffilmark{1,8} and
Jay Anderson\altaffilmark{9}}

\altaffiltext{1}{Department of Physics and Astronomy, San Francisco State University, 1600 Holloway Ave., San Francisco, CA 94132, USA; cool@sfsu.edu}
\altaffiltext{2}{Center for Interdisciplinary Exploration and Research in Astrophysics, Physics and Astronomy Department, Northwestern University, 2145 Sheridan Road, Evanston, IL 60208, USA; dhaggard@northwestern.edu}
\altaffiltext{3}{CIERA Postdoctoral Fellow}
\altaffiltext{4}{Dept. of Atmospheric and Oceanic Sciences, UCLA, Los Angeles, CA 90095, USA}
\altaffiltext{5}{Department of Physics, University of Washington, Seattle, WA 98195, USA}
\altaffiltext{6}{Bays Mountain Park \&\ Planetarium, Kingsport, TN 37660, USA}
\altaffiltext{7}{JATO Aviation, San Carlos, CA, 94070, USA}
\altaffiltext{8}{Astronomical Society of the Pacific, San Francisco, CA 94112, USA}
\altaffiltext{9}{Space Telescope Science Institute, Baltimore, MD 21218, USA}

\begin{abstract}

We present results of a search for optical counterparts of X-ray
sources in and toward the globular cluster Omega Centauri (NGC~5139)
using the Advanced Camera for Surveys (ACS) on the {\it Hubble Space 
Telescope}.  The ACS data consist of a mosaic of Wide Field Channel
(WFC) images obtained using F625W, F435W, and F658N filters; with 9
pointings we cover the central \about 10\amin \x 10\amin\ of the
cluster and encompass 109 known \Chandra\ sources.  We find promising
optical counterparts for 59 of the sources, \about 40 of which are
likely to be associated with the cluster.  These include 27 candidate
cataclysmic variables (CVs), 24 of which are reported here for the
first time.  Fourteen of the CV candidates are very faint, with
absolute magnitudes in the range \Mr\ $= 10.4-12.6$, making them
comparable in brightness to field CVs near the period minimum
discovered in the SDSS (G\"{a}nsicke et al.\ 2009).  Additional
optical counterparts include three BY~Dra candidates, a possible blue
straggler, and a previously-reported quiescent low-mass X-ray binary
(Haggard et al.\ 2004).  We also identify three foreground stars and
11 probable active galactic nuclei.  Finally, we report the discovery
of a group of seven stars whose X-ray properties are suggestive of
magnetically active binaries, and whose optical counterparts lie on or
very near the metal-rich anomalous giant and subgiant branches in
\wcen.  If the apparent association between these seven stars and the
RGB/SGB-a stars is real, then the frequency of X-ray sources in this
metal-rich population is enhanced by a factor of at least five
relative to the other giant and subgiant populations in the cluster.
If these stars are not members of the metal-rich population, then they
bring to 20 the total number of red stragglers (also known as
sub-subgiants) that have been identified in \wcen, the largest number
yet known in any globular cluster.

\end{abstract}


% SECTION 1

\section{Introduction}

As the most massive globular cluster in the Milky Way, and the first
to reveal the presence of multiple stellar populations, \wcen\ has
garnered significant attention from observers and theorists alike
(e.g., Cannon \&\ Stobie 1973, Freeman \&\ Rodgers 1975, Norris \&\ Da
Costa 1995, Norris, Freeman, \&\ Mighell 1996, Suntzeff \&\ Kraft
1996, van Leeuwen, Hughes, \&\ Piotto 2002).  Detailed studies of its
stellar populations reveal a remarkable and unexpected complexity,
beginning with the discovery of an anomalous red giant branch (Lee et
al.\ 1999, Pancino et al.\ 2000) with metallicity ten times that of
the majority population of giants in the cluster (Pancino et al.\
2000, Sollima et al.\ 2005, Johnson \&\ Pilachowski 2010).  Since
then, it has also been shown to harbor a double main sequence
(Anderson 1997, 2002a, 2003; Bedin et al.\ 2004), and a multiplicity
of subgiant branches (Ferraro et al.\ 2004; Villanova et al.\ 2007).
A large helium enhancement, possibly caused by early self-enrichment
from AGB stars (Renzini 2008, D'Ercole et al.\ 2008), has been invoked
(Norris 2004, King et al.\ 2012) to explain the startling finding that
the redder of the primary main-sequence populations is the more
metal-poor of the two (Piotto et al.\ 2005).  There is considerable
debate as to whether \wcen\ is a globular cluster at all.  Growing
evidence points instead to its being the stripped remnant of a dwarf
galaxy accreted by the Milky Way (Norris et al.\ 1996, Lee et al.\
1999, Bekki \&\ Freeman 2003, Renzini 2008).

A complete picture of the stellar populations in a globular cluster
includes its binary stars.  Like single stars, binaries can provide
insight into the conditions of a cluster's formation.  Knowledge of
binary populations is also central to an understanding of cluster
dynamical evolution (see, e.g., Hut et al.\ 1992).  With a half-mass
relaxation time approaching the Hubble time, and a moderate central
density of \about 1500 \lsun/pc$^3$ (Djorgovski 1993), many of \wcen's
primordial binaries are likely to have survived to the present day
(Davies 1997, Ivanova et al.\ 2006).  At the same time, its unusually
large core (r$_c$ $\simeq$ 3.7 pc) is such that significant numbers of
stellar collisions and near misses are expected to have occurred over
its lifetime (Verbunt \&\ Meylan 1988, Di Stefano \&\ Rappaport 1994).
Regardless of its origins, the sheer number of binaries that \wcen\ is
likely to harbor by virtue of its enormous mass (\about 3 \x\
$10{^6}$\msun; Meylan 2002) provides an opportunity to observe large
numbers of binaries all at essentially the same distance, and to
uncover potentially rare classes of systems.

One fruitful way to search for binary stars in globular clusters is
via X-ray imaging.  At the limiting luminosities reached in nearby
globular clusters with the {\it Chandra X-ray Observatory}, a diverse
array of binaries can be revealed.  Beyond the high-luminosity
globular cluster sources known since the early days of X-ray astronomy
(Giacconi et al.\ 1974) and understood to be accreting neutron stars
(Clark 1975), the much more abundant low-luminosity X-ray sources
include cataclysmic variables (CVs; Hertz \&\ Grindlay 1983a),
quiescent low-mass X-ray binaries (qLMXBs; Verbunt, van Paradijs, \&\
Elson 1984), millisecond pulsars (MSPs; e.g., Grindlay et al.\ 2002)
and chromospherically active stars.  The latter may be the result of
enhanced coronal activity due to tidal locking in a binary system;
hereafter we shall use the term ``active binary'' (AB) to refer to
either the main-sequence variety (BY~Dra stars---Dempsey et al.\ 1997)
or the subgiant variety (RS~CVn stars---Dempsey et al.\ 1993), both of
which have been seen in globular clusters (e.g., Kaluzny et al.\ 1996,
Taylor et al.\ 2001, Albrow et al.\ 2001).  Among the low-luminosity
sources, only the qLMXBs (also known as quiscent neutron stars, or
qNS), with their distinctive soft X-ray spectra and moderate
luminosities, can be identified on the basis of X-ray observations
alone (Brown, Bildsten, \&\ Rutledge 1998, Rutledge et al.\ 2000).
For others, optical (or radio, in the case of MSPs) follow-up is
essential.  In the optical, the resolving power of the {\it Hubble
Space Telescope} (\HST) is crucial; the vast majority of CVs now known
in globulars are hopelessly lost in the light of brighter neighbors in
ground-based imaging.

Early X-ray imaging of \wcen\ with Einstein IPC revealed multiple
low-luminosity X-ray sources in and toward the cluster (Hertz \&\
Grindlay 1983b).  The source closest to the cluster center, IPC source
C, was subsequently resolved into three separate sources with ROSAT
HRI (Verbunt \&\ Johnston 2000).  Using the HRI positions, Carson,
Cool, \&\ Grindlay (2000) identified two of these as probable CVs
based on the discovery of \ha-bright, UV-bright optical counterparts
in \HST/WFPC2 imaging.  IPC source B, \about 4\amin\ from the cluster
center, coincides with a \Chandra\ source subsequently identified as a
qNS (Rutledge et al.\ 2002, Haggard et al.\ 2004).  Two other sources
(``A'' and ``D''), both more than 10\amin\ from the cluster center,
were shown to be foreground dMe stars (Cool et al.\ 1995a).  Gendre et
al.\ (2003) increased to 27 the total number of X-ray sources known
within the half-mass radius of \wcen\ using {\it XMM-Newton}, and found
that their X-ray properties are consistent with their being a
combination of CVs and active binaries.  More recently, a 69 ksec
ACIS-I exposure with \Chandra\ that reached a limiting luminosity of
\lx \about 10$^{30}$ \ergs\ revealed 180 X-ray sources in and toward
the cluster, 81 within the half-mass radius (Haggard, Cool, \&\ Davies
2009, hereafter HCD09).

One of the challenges in identifying binary stars in \wcen\ using
X-ray imaging is its large angular size on the sky.  Owing to this, we
expect a significant fraction of the \Chandra\ sources to be active
galaxies behind the cluster (AGN; see HCD09 for a detailed
discussion).  With the exception of the qNS, no unique X-ray spectral
signature allows us to distinguish probable AGN from accreting
binaries in the cluster, particularly given the faintness of the
sources.  To determine which of these sources are binaries within the
cluster, we must therefore search for optical counterparts for all the
X-ray sources present, with the expectation that only a fraction are
likely to be associated with the cluster.

Here we report the results of a search for optical counterparts of
\Chandra\ sources using \HST's Advanced Camera for Surveys (ACS).
Preliminary results of this study were reported by Haggard et al.\
(2010).  The ACS/WFC data consist of a mosaic of 3 \x\ 3 pointings
which encompasses 109 of the \Chandra\ sources.  We use the \ha\
imaging technique that has been applied successfully to searches for
CVs in other clusters (e.g., Cool et al.\ 1995b, Grindlay et al.\
1995, Bailyn et al.\ 1996, Pooley et al.\ 2002, Anderson, Cool, \&\
King 2003, Cohn et al.\ 2010).  This method is also sensitive to
qLMXBs which, like CVs, are characterized by strong hydrogen emission
lines.  Using these data, an optical counterpart for the source
initially identified as a qNS on the basis of its X-ray spectrum alone
(Rutledge et al.\ 2002) was reported by Haggard et al.\ (2004).  While
chromospherically active binaries have rather weak emission lines
compared to accreting binaries (e.g., Chevalier \&\ Ilovaisky 1997),
the subset with the greatest coronal activity can also be identified
using this \ha\ imaging technique (e.g., Taylor et al.\ 2001, Cohn et
al.\ 2010).

The outline of the paper is as follows.  In Section 2 we describe the
observations, together with the astrometric and photometric techniques
used to analyze the data.  We then describe in detail the method by
which we evaluated potential optical counterparts.  In Section 3 we
present the most promising optical counterpart identifications,
dividing them into categories based on their locations in \br\ vs.\
\r\ and \hr\ vs.\ \r\ color--magnitude diagrams (CMDs).  We discuss
the results in Section 4 and summarize them in Section 5.

% SECTION 2

\begin{figure*}
\plotone{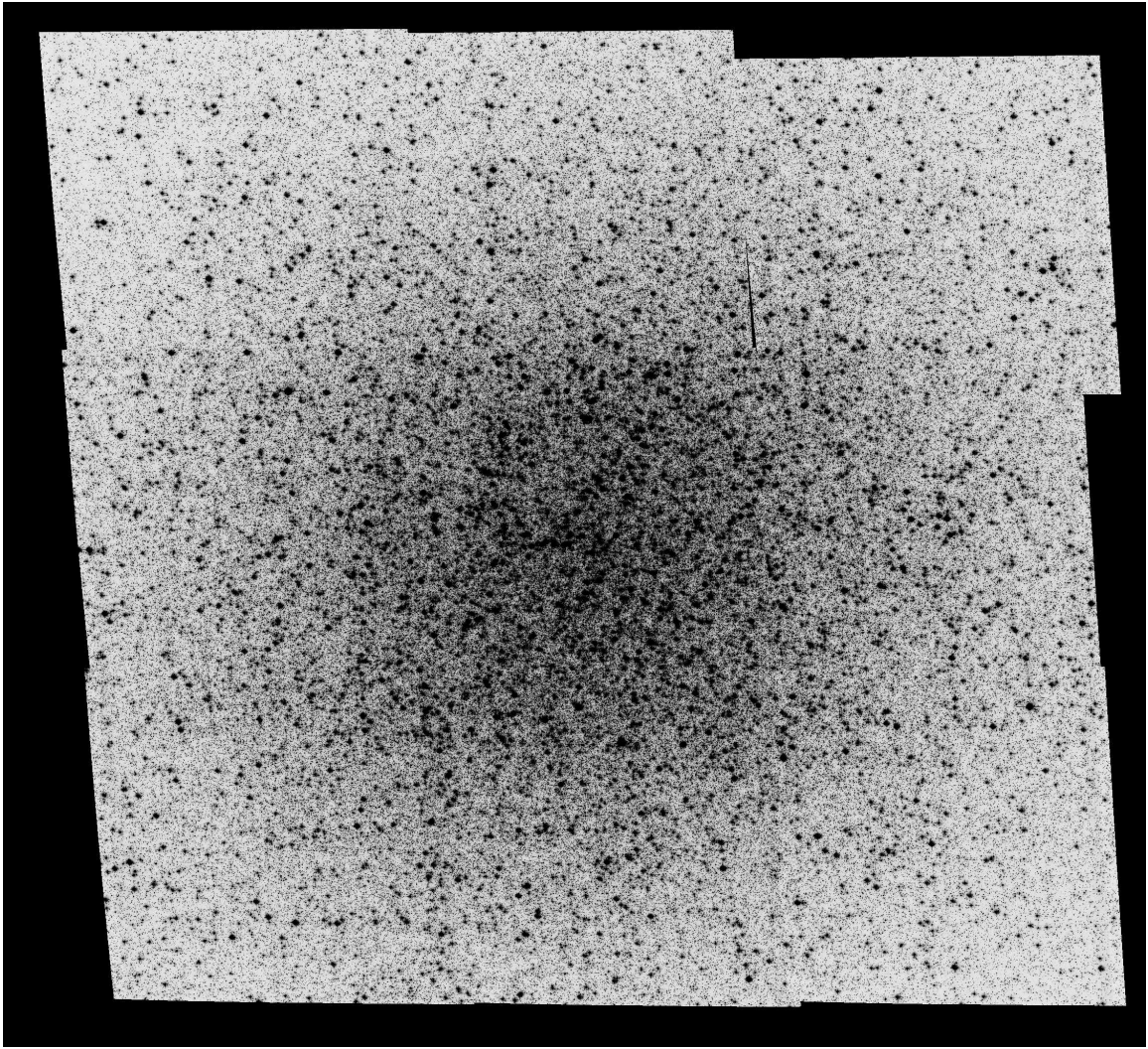}
\figcaption{Mosaic of \wcen\ constructed from ACS/WFC
\b\ images at nine different pointings.  Note that the northwest
tile was slightly offset from the planned pointing, apparently due to the
misaquisition of guide stars.  The field of view is approximately 10\amin\ \x\ 10\amin; north is up and east is to the left.}
\end{figure*}

\section{Observations and Analysis}

We obtained images of \wcen\ with the Wide Field Channel (WFC) of the
Advanced Camera for Surveys (ACS) on \HST\ using F625W (R$_{625}$),
F435W (B$_{435}$), and F658N (\ha) filters on 2002 June $27-30$.  The
observations consist of a 3 \x\ 3 mosaic covering \about 10\amin\ \x
10\amin\ approximately centered on the cluster center (Fig.\ 1).  The
full mosaic extends beyond the cluster's half mass radius of \about
4.2\amin\ (Harris 1996) and emcompasses 109 of the 180 X-ray sources
we detected in \Chandra\ observations of \wcen\ (HCD09).

At each of the 9 tiles in the mosaic, we obtained a total of 12
exposures: 4 \x\ 440 s with the \ha\ filter (F658N), 3 \x\ 340 s with
each of the broad-band filters (F625W and F435W), and one short
exposure in each of the broad-band filters (8s and 12s in F625W and
F435W, respectively).  A gain of 2 electrons/DN was adopted so as to
make full use of the \about 85,000 electron well-depth of the WFC
CCDs.  Shifts of \about 5\asec\ were made between each of the 4 \ha\
exposures.  In the two broad-band filters, shifts of 7\secspt 5 were
adopted so that the 3 available exposures span the same region as the
\ha\ exposures.  These ``dither'' patterns insure that any given star
in the mosaic lands in the \about 2\secspt 5-wide chip gap at most
once in any filter.  With a minimum of 3 long exposures in each filter
we are also able to identify cosmic rays reliably.

% SECTION 2.1

\subsection{Astrometry}

To map the positions of \Chandra\ sources onto the ACS/WFC images, we
first applied a distortion correction to the individual WFC images
using the solution obtained for the F475W filter from a study of
47~Tuc (Anderson 2002b).  The internal accuracy of this solution
should be \about\ 0.15 WFC pixel ($\simeq$ 0\secspt 0075).  We then
stitched together all the individual \b\ frames to make a mosaic of
the entire field in that filter.  The images fit together well, with
no sign of misaligned star images where chips overlap, indicating that
the distortion correction was working well.

To determine the R.A. and Dec. associated with stars on the mosaic, we
used the star lists of Kaluzny et al.\ (1996) and van Leeuwen et al.\
(2000).  More than 15,000 of the former and 4000 of the latter fall
within the field of view of our mosaic.  There is a small zeropoint
offset of 0\secspt 5 between the two systems, which we take to be
indicative of the uncertainty in the absolute frame; we used the
Kaluzny system to define the transformation between our mosaic system
and R.A. and Dec.  We estimate that, at any given location in the
mosaic, the absolute astrometry should be accurate to \simlt 0\secspt
2 due to uncertainties in stitching together the individual images.

In our search for optical counterparts we intially adopted 1\secspt 0
error circles.  This radius was chosen in consideration of the
0\secspt 6 uncertainty (90\% confidence) in \Chandra's absolute
coordinate system, uncertainties in $wavdetect$ positions at the level
of \about 0\secspt 5 several arcminutes off-axis (Feigelson et al.\
2002), and the \simlt 0\secspt 2 uncertainties in our mosaic
construction.  As a check on whether this error budget was
sufficiently generous, we recovered the two stars identified by Carson
et al.\ (2000) as probable CVs; both are $<$0\secspt 3 from their
nominal positions.  The optical counterpart to the qNS identified by
Haggard et al.\ (2004), 4\minspt 4 from the cluster center, is
$<$0\secspt 5 from the nominal position and thus also well within the
1\asec\ circle.

After we had analyzed roughly one third of the sources we computed a
boresight correction using 14 promising optical counterparts
identified as of then, including the 4 objects found in our previous
studies of \wcen\ (Carson et al.\ 2000, Haggard et al.\ 2004).  The
original error circle was centered at pixel coordinate (200.5, 200.5)
on the 401 \x\ 401 pixel ``patches'' that we extracted surrounding
each X-ray source (see \S 2.2) and had a radius of 20 pixels (1\secspt
0).  The boresight-corrected error circle is centered at (206.1,
206.6).  The standard deviation of the offsets between the observed
positions of these 14 optical counterparts and this new center was
\about 4 pixels in both x and y.  Based on this, we chose a 12 pixel
(0\secspt 6) radius (\about 3$\sigma$) for the new error
circle.\footnote {For a bivariate normal distribution, this is
  equivalent to a 99\% error circle.}  This new circle lies almost
entirely within the larger 1\asec\ circle (see Fig.\ 4), but has an
area only 0.36 times as large (see also \S 2.3).  For simplicity we
have adopted the same size error circle for all sources; we verify
below (see \S 3.2) that it is sufficiently generous even for faint
X-ray sources for which the uncertainty in the X-ray position
dominates the error budget.  The full set of optical identifications
appears in Table~1.

% SECTION 2.2

\subsection{Photometry}

We began with ``flt'' images that had been processed through the
standard \HST\ calibration pipeline, including debiasing, dark
subtraction, and flat-field correction, using the best available
calibration images as of 2004 August 05.  We then used the data
quality files to identify saturated pixels in each image and set them
to a high value (99000) so that they would be easily recognizable in
the subsequent analysis.  We extracted 401 \x\ 401 pixel (\about
20\asec \x\ 20\asec) ``patches'' around each of the \Chandra\ source
positions (HCD09) in each of the images.\footnote{Patch images may be
found at http://www.physics.sfsu.edu/$\sim$cool/omegaCen/.}  For 95
of the 109 sources, the result is a set of 12 small images to be
analyzed, with the source position close to the center of each.  For
13 of the sources, the fact that the tiles of the mosaic overlap one
another meant that images were available from more than one tile.  In
these cases, we analyzed images from the tile with the most images,
which was a full set of 12 in all but two cases (24d with 9 images,
and 24f with 8 images).  One source (31d) falls near a small gap
created by the slightly mis-pointed tile (see Fig.\ 1), but still has
coverage in 9 images.

Despite the high resolution of the ACS/WFC (0\secspt 05 pixels), the
images are still quite crowded, with many stars overlapping one
another.  We therefore chose to use DAOPHOT and ALLSTAR (Stetson 1987)
for the analysis.  We also found that DAOPHOT/ALLFRAME (Stetson 1994),
with its capacity to analyze multiple images simultaneously and
require consistent positions for all stars, was very valuable given
the level of crowding and numerous cosmic rays in the images.  The two
labor-intensive parts of the analysis are the creation of high-quality
point-spread functions (PSFs) and the production of reliable and
complete star lists in the region of the X-ray source error circles.
Because of the possibility that an interesting optical counterpart
could be missed in any fully automated process, we decided to take a
more hands-on approach.  We describe each of these steps in detail
below.

We began by creating PSFs empirically for each individual long
exposure of each patch.  For the short \r\ and \b\ exposures, the PSF
from the long exposure at the corresponding dithers were adopted.  Our
earlier tests on the patch containing the qLMXB (Haggard et al.\ 2004)
had shown that separate PSFs for each exposure produced the best
results.  We found that, when analyzing these small image patches,
constant PSFs worked well.  Typically \about $10-40$ of the most
isolated stars were used to define the PSF in each case, including a
few saturated stars to help to better define the model at large radii.
PSF radii of $7-11$ pixels were adopted, with the smaller radii
generally being required in the most crowded patches.  These PSF sizes
were adequate to model and remove most stars in the X-ray error
circles very well.  We iterated the PSF-creation process as many times
as needed, removing neighbor stars, checking the quality of
subtractions, and removing and/or adding more PSF stars as needed.

After extensive experimentation with DAOPHOT and PSF-fitting
techniques on the WFC ``flt'' images, we found that pixels adjacent to
saturated pixels were also affected by saturation.  This became
apparent when we found that the measured magnitudes of saturated stars
in long exposures were systematically brighter than those obtained
from the short exposures.  We therefore decided to flag all pixels
adjacent to saturated pixels and ignore them in both the creation of
PSFs and the generation of magnitudes.  This eliminated the systematic
bias in the measurement of saturated stars.  The quality of PSFs also
improved substantially, allowing us to make much better use of
saturated stars to determine PSF structures at large radii.

Initial star lists were obtained using automated DAOPHOT/FIND on one
of the \r\ images.  For each patch, we then examined by eye the region
surrounding the \Chandra\ source position at the center of the patch.
We focused initially on a 1\asec\ error circle surrounding the nominal
X-ray source position, and later narrowed our search to a 0\secspt 6
circle once the boresight correction had been applied (see \S 2.1).
We began by removing from the list any objects that were clearly
cosmic rays or other artifacts, which we determined by blinking the
different images against one another.  We also adopted a conservative
approach and removed any objects that we were not confident were real
(e.g., in regions where multiple stars overlapped).  We then ran
ALLSTAR and ALLFRAME and carefully examined the subtracted images.
Remaining objects that appeared consistently in all 3 subtracted \r\
and/or \b\ images were then added to the star list by hand.  This
procedure was repeated until no additional objects could be seen in
the subtracted images.  Typically 3 to 5 iterations were needed to
produce the final list of stars.  This work was done primarily using
the 3 long \r\ exposures, which include the faintest stars.  We also
examined the \ha\ and long \b\ exposures to check for any additional
very blue or \ha-bright objects.

When the PSFs and star lists were finalized we ran ALLSTAR and
ALLFRAME one last time to generate final positions and magnitudes for
all the stars.  We then matched stars up across the 12 frames,
requiring that a star be found in at least 4 of the frames for it to
be included in the final list.  We retained both ALLSTAR and ALLFRAME
results because, in a few cases, the distortion of the ACS/WFC was
sufficiently large (e.g., when a source lands near an edge in one
dither) that the subtracted images in ALLFRAME show evidence of
imperfect star positions.  In these cases we carefully assessed
whether ALLSTAR or ALLFRAME gave more reliable results by examining
subtracted images and the width of the main sequence in
color-magnitude diagrams (CMDs).  In the one case (21c) where the
ALLSTAR results were deemed more reliable (and a candidate optical
counterpart was found), we report these results in Fig.\ 3 and Table
1.  We transformed our instrumental magnitudes to the VEGAMAG system
using the zeropoints and aperture corrections provided by Sirianni et
al.\ (2005) for the ACS/WFC.

% SECTION 2.3

\begin{figure}
%\plotone{f2.eps} 
\centerline{
\includegraphics[scale=0.45]{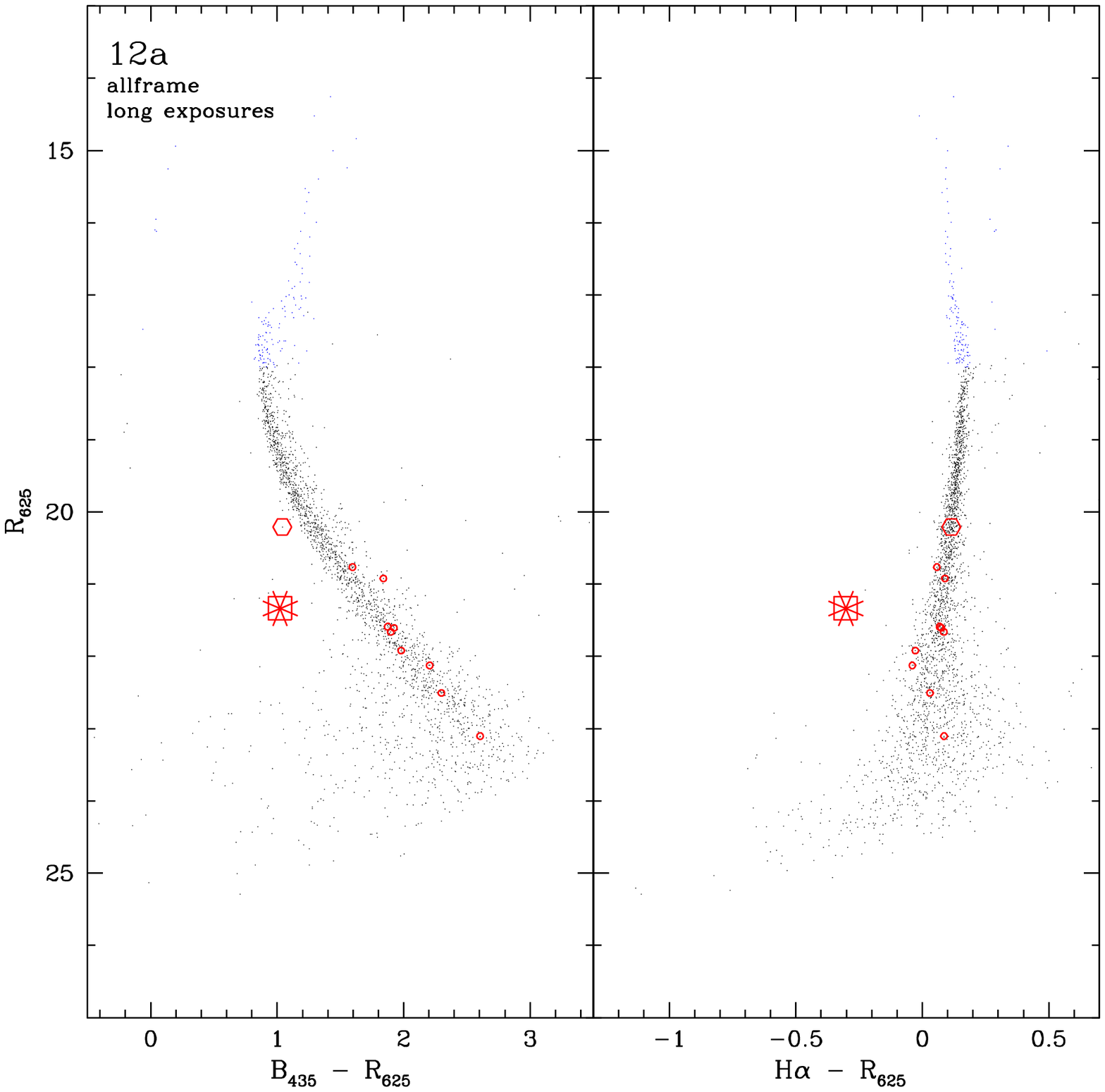}
}
\figcaption{\br\ and \hr\ color-magnitude diagrams for stars
in the vicinity of \Chandra\ X-ray source 12a.  Small dots are all
stars in a 20\asec\ \x\ 20\asec\ patch surrounding the source position
(saturated stars in blue, non-saturated stars in black).  Stars that lie inside the
0.6\asec\ error circle that was scrutinized in detail are marked with
open red symbols.  Small red circles are stars that lie on the main
sequence or SGB or RGB in both the \br\ vs.\ \r\ and
\hr\ vs.\ \r\ CMDs.  Larger symbols are used to indicate objects that
lie off these sequences in one or both diagrams.  Symbol shapes
indicate the quality of the candidates (see \S 2.4).  An asterisk
marks the object most likely to be the optical counterpart of the
X-ray source.  In this case it is an object that is both blue and
\ha-bright: a probable cataclysmic variable (see \S 3.1).}
\end{figure}

\subsection{Candidate Inspection}

Once we had constructed CMDs for each patch, we carried out a
systematic evaluation of all potentially interesting objects.  All
objects that did not lie on or very near the main sequence or giant
branch in both the \br\ and \hr\ CMDs were evaluated for reliability
of the photometry.  We did this by examining the object in all the
individual images to check the potential impact of near neighbors,
cosmic rays, and diffraction spikes, and by looking at how cleanly
DAOPHOT removed it from each of the images.  We also took account of
the consistency of the multiple independent measurements in the
different filters.  During this part of the analysis we carefully
examined all the subtracted images to look for signs that any of the
objects were extended (i.e., whether flux was left behind after a
point source was subtracted), and for additional objects that might
have been missed in previous rounds.  When additional objects were
found, we added them to the list and remade the CMDs before further
evaluation.

Based on these considerations, objects of potential interest were
classified as having one of four ``qualities,'' 0 through 3.  Objects
for which no potential problems were found, such that their positions
in the CMDs are very reliable, were classified as having quality 0.
Objects for which only relatively minor problems were present, such
that their positions in the CMDs should still be reliable, were
classified as quality 1.  Objects for which potentially significant
problems were found, such that their positions in the CMD might not be
reliable, were classified as quality 2.  Objects that were apparently
real stars, but for which significant problems were found
(e.g. diffraction spike, much brighter neighbor very close by, etc.),
such that their positions in the CMDs are unlikely to be reliable,
were classified as quality 3.  Here it should be made clear that the
quality flag is an assessment of the reliability of an object's
measured position within the CMDs (i.e., whether the object can be
reliably said to be on or away from the nearest well-populated
sequence), not of the probability of its being the actual optical
counterpart of the X-ray source.

As an example, we show in Figure~2 the CMDs for the patch
corresponding to the northernmost of the three bright X-ray sources in
the core (\Chandra\ source 12a, HCD09), referred to by Carson et al.\
(2000) as XC.  The small dots in the image are all stars within the
\about 20\asec\ \x\ 20\asec\ patch; these include many artifacts and
poorly measured stars since no effort was made to ``clean'' the area
of the patch outside the X-ray error circle.  Small circles represent
stars that lie within the 0\secspt 6 error circle; these objects
should all be real, as they were carefully assessed by eye.  Larger
symbols are used to represent those objects that were scrutinized for
photometric reliability; different symbols are used to represent
different qualities (triangle, square, pentagon and hexagon for
quality 0, 1, 2, and 3, respectively).  In this particular patch, 11
objects landed inside the error circle.  Two lie off the main sequence
in one or both CMDs and were assessed for reliability.  Of these, one
was deemed likely to be reliable (the square) while the other was
deemed likely not to be (the hexagon).  Finally, we added an asterisk
to the object that we deemed most likely to represent the optical
counterpart of the X-ray source.  Only quality 0 and 1 objects were
considered as potential counterparts.  In this patch there was one
such object.  It is both blue and \ha-bright; we show below (see \S
3.1) that it is likely to be a cataclysmic variable, a classification
supported by previous work (HCD09 and references therein).

%\begin{figure*}
%\centering
%  \subfigure[]{\includegraphics[scale=0.85]{f3a.eps}}
%  \subfigure[]{\includegraphics[scale=0.85]{f3b.eps}}
%  \subfigure[]{\includegraphics[scale=0.85]{f3c.eps}}
%  \subfigure[]{\includegraphics[scale=0.85]{f3d.eps} } 
%\figcaption{Full set of 59 CMD pairs; symbols as in Fig. 2.}
%\end{figure*}

\begin{figure*}
\centerline{\includegraphics[scale=0.83]{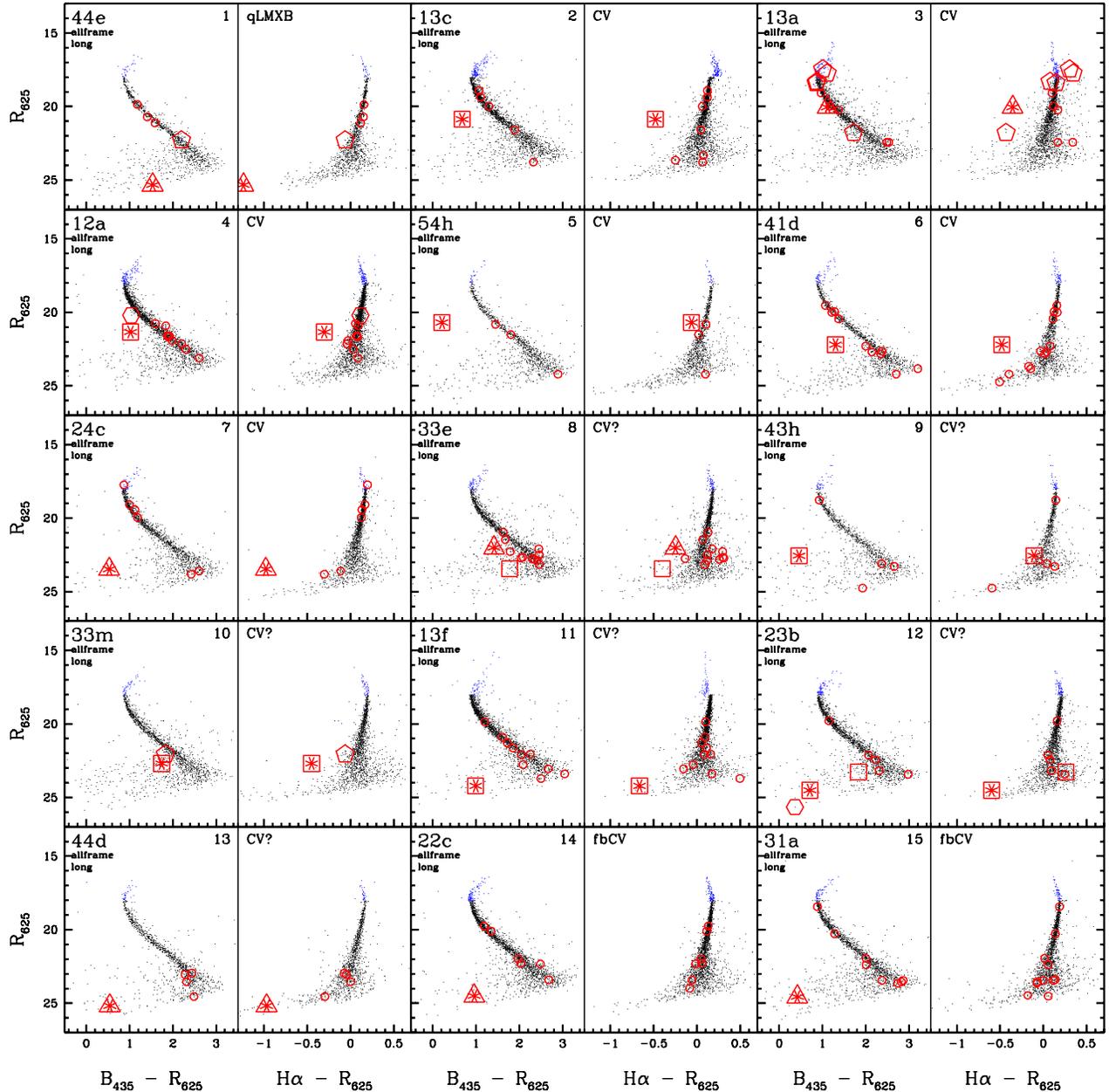}}
%\centering
%\subfloat[Full set of 59 CMD pairs; symbols as in Fig. 2.]{\includegraphics[scale=0.85]{f3a.eps}}
\centerline{\figcaption[]{Full set of 59 CMD pairs; symbols as in Fig. 2.}}
\end{figure*}

\begin{figure*}
\ContinuedFloat
\centerline{\includegraphics[scale=0.83]{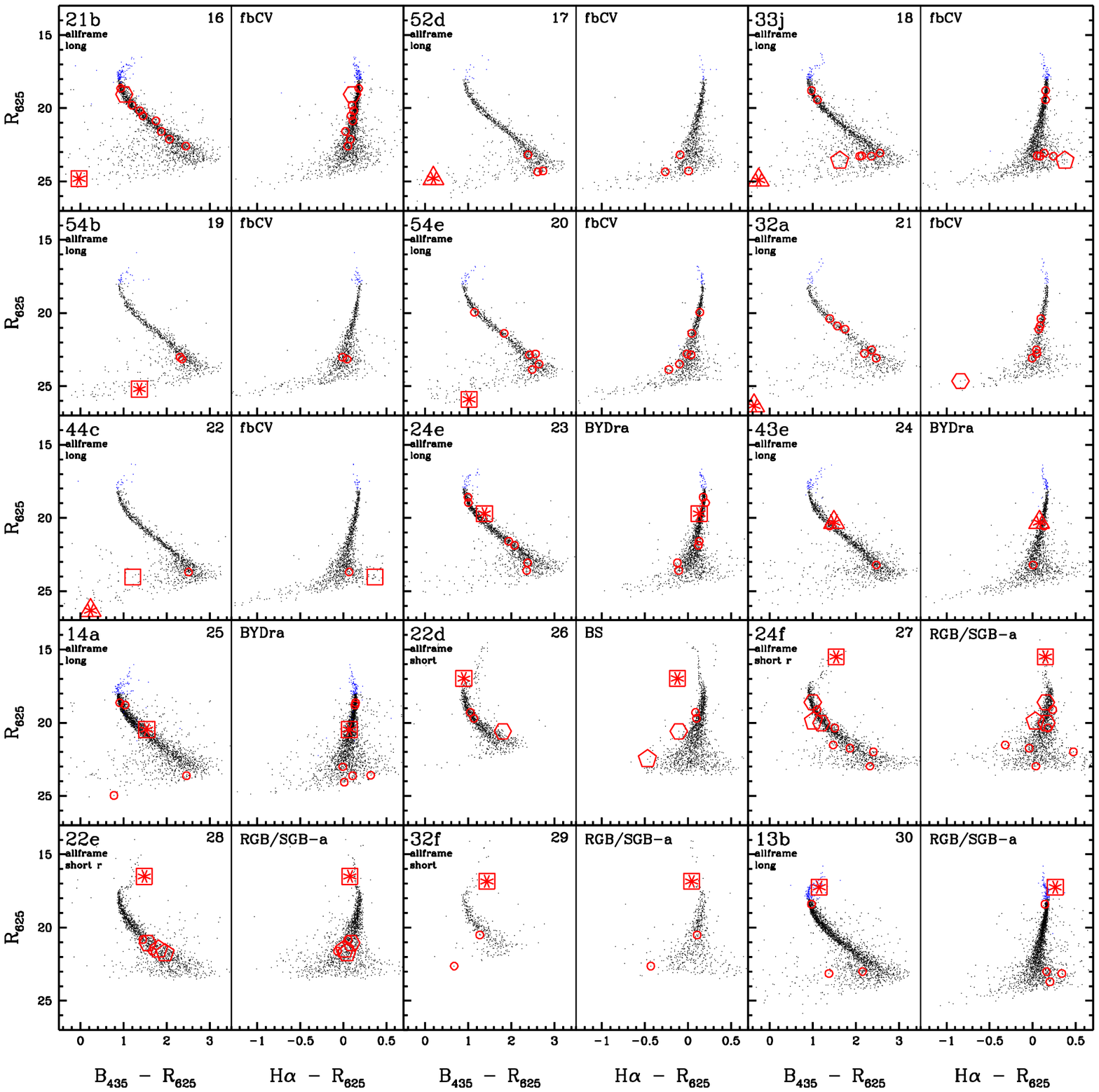}}
\figcaption[]{continued}
\end{figure*}

\begin{figure*}
\ContinuedFloat
\centerline{\includegraphics[scale=0.83]{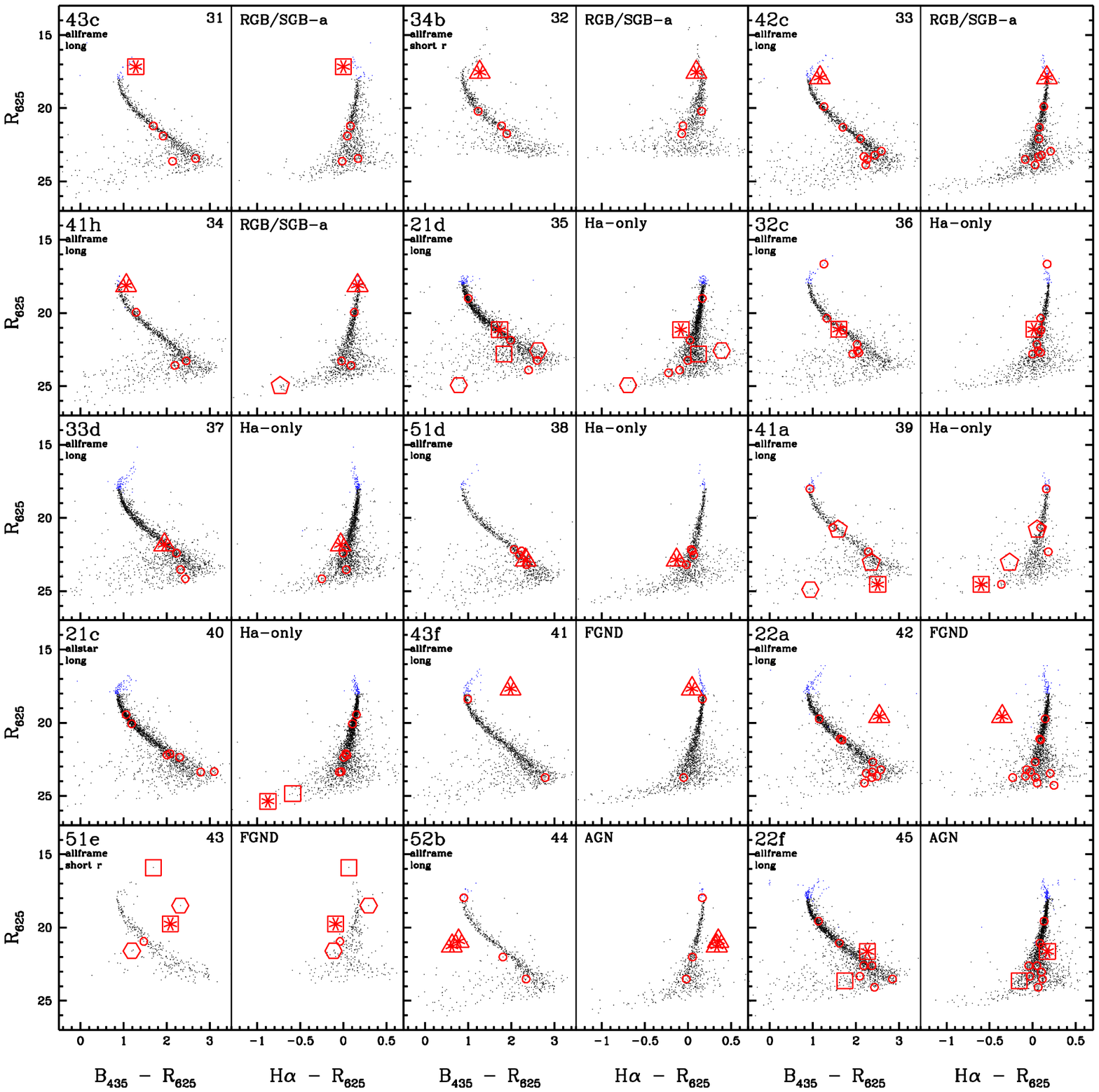}}
\figcaption[]{continued}
\end{figure*}

\begin{figure*}
\ContinuedFloat
\centerline{\includegraphics[scale=0.83]{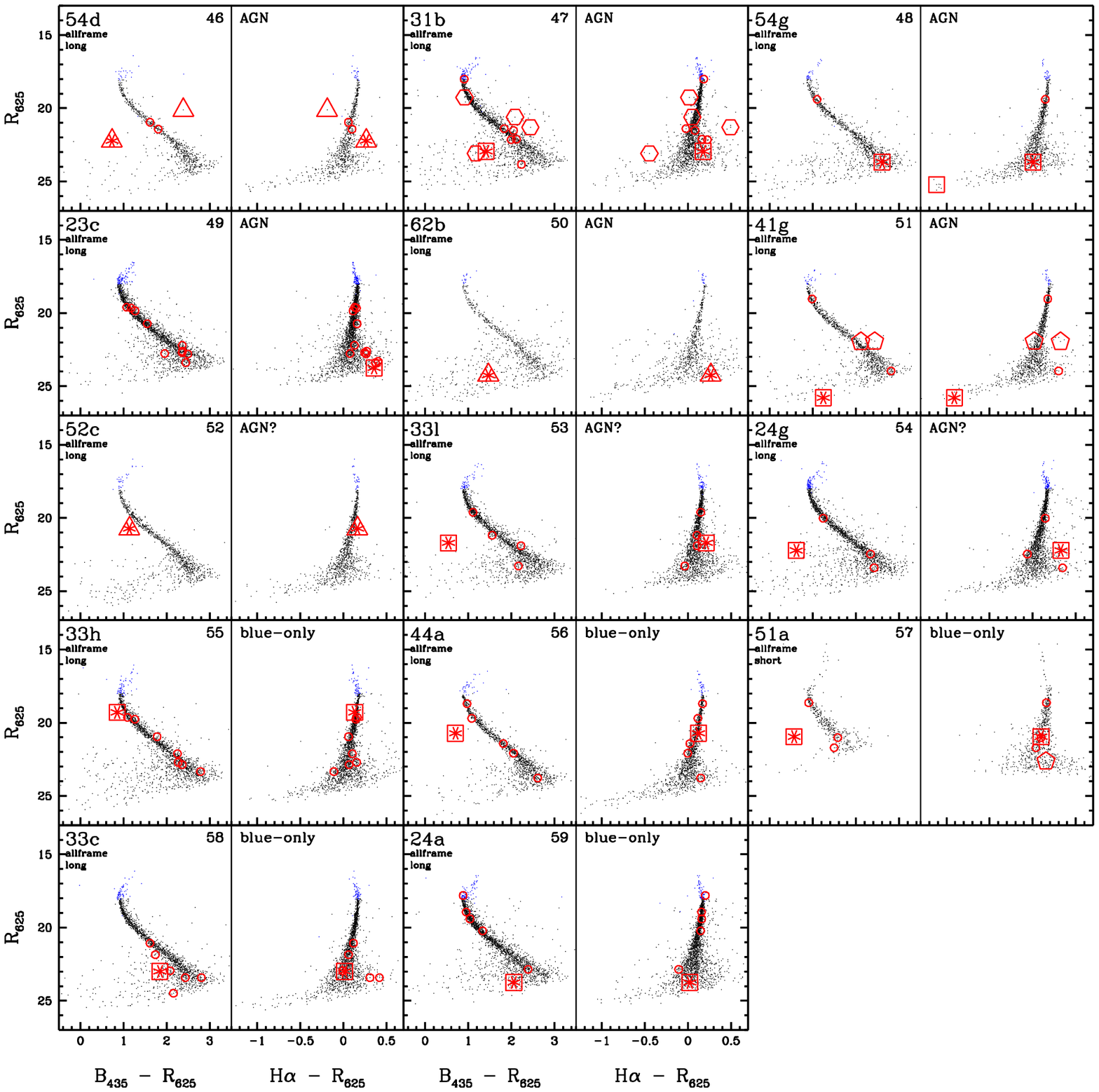}}
\figcaption[]{continued}
\end{figure*}

In Fig.\ 3 we present CMDs like those in Fig.\ 2 for the 59 patches in
which one or more quality 0 or 1 object was found to lie off the main
sequence in one or both diagrams (\br\ vs.\ \r\ and/or \hr\ vs.\
\r).\footnote{Larger versions of these CMDs may be found at 
http://www.physics.sfsu.edu/$\sim$cool/omegaCen/.} Each pair of
panels corresponds to one X-ray source, with the name of the source
from HCD09 listed in the upper left corner of the left panel.  For
ease of comparing related candidates to one another, CMDs are grouped
according to the classification of the candidate (see \S 3) and sorted
in order of \r\ magnitude within each class.  Panel pairs are numbered
1 to 59 in the upper right corner of the left-hand panel in each pair.
For each pair of panels we also list the type of photometry presented
(ALLFRAME or ALLSTAR), and the type of exposures from which the CMDs
were derived (``long,'' ``short'' or ``short r'').  In most cases we
present the photometry derived from the long \r\ and \b\ exposures
(``long''); in \ha\ there are no short exposures in any case.  In some
cases, however, either the candidate itself or a near neighbor was
saturated in the long \r\ exposures and/or long \b\ exposures.  In
cases where saturation affected the reliability of the photometry in
long exposures we present CMDs derived from either the short \r\ and
long \b\ exposures (``short r'') or short \r\ and short \b\ exposures
(``short''), depending on whether only the long \r\ exposures were
affected or both the long \r\ and long \b\ exposures.  Finally, in the
upper left corner of the right-hand panel in each pair we list a
category for each of the optical counterparts identified---see \S 3.
In some cases these designations are necessarily tentative.

In Table 1 we list all quality 0 and 1 candidate optical counterparts
found in these 59 patches.  Column 1 lists the X-ray source ID from
HCD09.  Columns 2, 3, and 4 are the ID number, and x and y coordinates
from the ALLFRAME (or ALLSTAR) reductions of the corresponding 401 \x\
401 pixel patch.\footnote{Coordinates correspond to the ``r1'' patch
image if available and otherwise to the ``r3'' image.}  The radial
offsets in pixels from the boresight-corrected \Chandra\ positions are
given in column 5, and the quality of the candidates (0 or 1) appear
in column 6.  Column 7 gives the classification assigned to each
counterpart in this paper.

In 49 of the patches, a single quality 0 or 1 object was found; it is
marked with an asterisk in Fig.\ 3 as the most probable optical
counterpart of the X-ray source.  In 10 cases, two quality 0 or 1
objects were found in the error circle (see Fig.\ 3).  In these cases,
the two potential counterparts are listed together in Table 1, with
the most probable ID listed first.  In all but 3 cases we choose the
object closest to the center of the error circle as the more probable
counterpart and mark it with an asterisk.  The exceptions are 21d,
22f, and 23b, for which the candidate closest to the center was a blue
star which was on the main sequence in \hr; such stars have a
relatively high liklihood of being chance coincidences (see \S 3.8).
For 23b we consider an object that is both blue and \ha-bright (a
possible CV) to be the more probable counterpart; for 21d, we consider
an \ha-bright star the more probable counterpart; and for 22f we
consider an extended object to be the more probable counterpart (a
likely AGN).  These objects are marked with an asterisk in Fig.\ 3.

We have verified that in excluding quality 2 objects from
consideration we are not missing significant numbers of objects that
would be promising optical counterparts.  Of the 59 sources for which
we report identifications, 13 error circles include one or more
quality 2 objects.  In only four cases (13a, 33m, 41g, and 41h) is a
quality 2 object closer to the center of the error circle than the
counterpart we report in Table 1.  A close inspection of the results
in each of these cases confirms that the quality 2 candidates in
question are likely to lie off the principal sequences as a result of
poor measurements and as such are unlikely to be the true counterparts
of the X-ray sources.

J2000 coordinates for all of the candidates appear in columns 8 and 9
of Table 1.  Offsets from the cluster center determined by Anderson
\&\ van der Marel (2010), R.A. = 13:26:47.24, Dec. = $−$47:28:46.45
(J2000) appear in column 10.  The coordinates provide here are tied to
the mosaic image created by Anderson \&\ van der Marel (2010) and can
thus be used to locate the stars on the mosaic image provided with
that paper.  We note that the mosaic was constructed using updated
distortion corrections relative to what had been used earlier to
locate the centers of the patches for the present work.  In most cases
the differences are negligible, though they can be as large as \about
0\secspt 2.  In a small number of cases this puts the candidate
optical counterpart slightly outside a 0\secspt 6 error circle, but
still well within a 1\asec\ circle.

The small changes in X-ray source positions resulting from the new
distortion correction do not alter our choices of which objects are
the most likely counterparts in the 10 cases where there are two
objects to choose from.  In six of those cases, the star that is
closest to the center of the 0\secspt 6 error circle in the patch is
also closest to the newly determined position on the mosaic image.  In
two of the three cases where we had selected an object that was
farther from the center of the error circle as the most likely
counterpart (22f and 23b), the chosen object turns out to be closer to
the new position, lending support to our choice.  In only two cases
(21c and 51e) does the star that had been closer to the center of the
error circle now lie farther away from the newly determined position.
In both these cases, the difference in offsets from the new position
is small (\about 0\secspt 1), such that position alone is not a very
useful discriminant.  In the case of 51e, where both potential
counterparts are probable foreground stars, we consider the original
choice to be the best one.  This star shows clear emission in \ha\
(see panel 43 in Fig.\ 3), which is strongly associated with emission
in X-rays.  In the case of 21c, both potential counterparts are in the
``\ha-only'' category and lie in a similar location in the \hr\ CMD
(see panel 40 in Fig.\ 3).  As there is little to distinguish between
the two, we choose not to alter our original selection.

In Table 2 we provide photometric information about each candidate
counterpart along with its tentative classification.  In the cases for
which two potential counterparts were identified, the alternate
countparts are listed separately at the end of the Table.  Entries are
sorted by classification and then by \r\ magnitude, to match the
sequence in which CMDs appear in Fig.\ 3.  The X-ray ID, \r\
magnitude, \br\ color, and \hr\ color appear in columns $1-4$,
respectively.  Column 5 indicates whether the object appeared blue or
red relative to the main sequence (or giant branch).  Column 6
indicates whether the object appears ``\ha-bright'' (to the left of
the main sequence) or ``\ha-faint'' (to the right).  Objects that land
on the main sequence (or giant branch) are listed as ``neither'' in
these columns.  Columns 7 and 8 list the quality and classification of
the counterparts, and column 9 lists the X-ray--to--optical flux
ratio.\footnote{X-ray fluxes from HCD09 have been revised downward by
a factor of 1.25 (see \S 4.1).}  For sources likely to be associated
with the cluster, column 10 gives the X-ray luminosity for an assumed
distance of 4.9 kpc and hydrogen column density of 9 \x\ 10$^{20}$
cm$^{-2}$.

\begin{figure}
\hspace{-1cm}
%\plotone{f4.eps}
\centerline{
\includegraphics[scale=0.7]{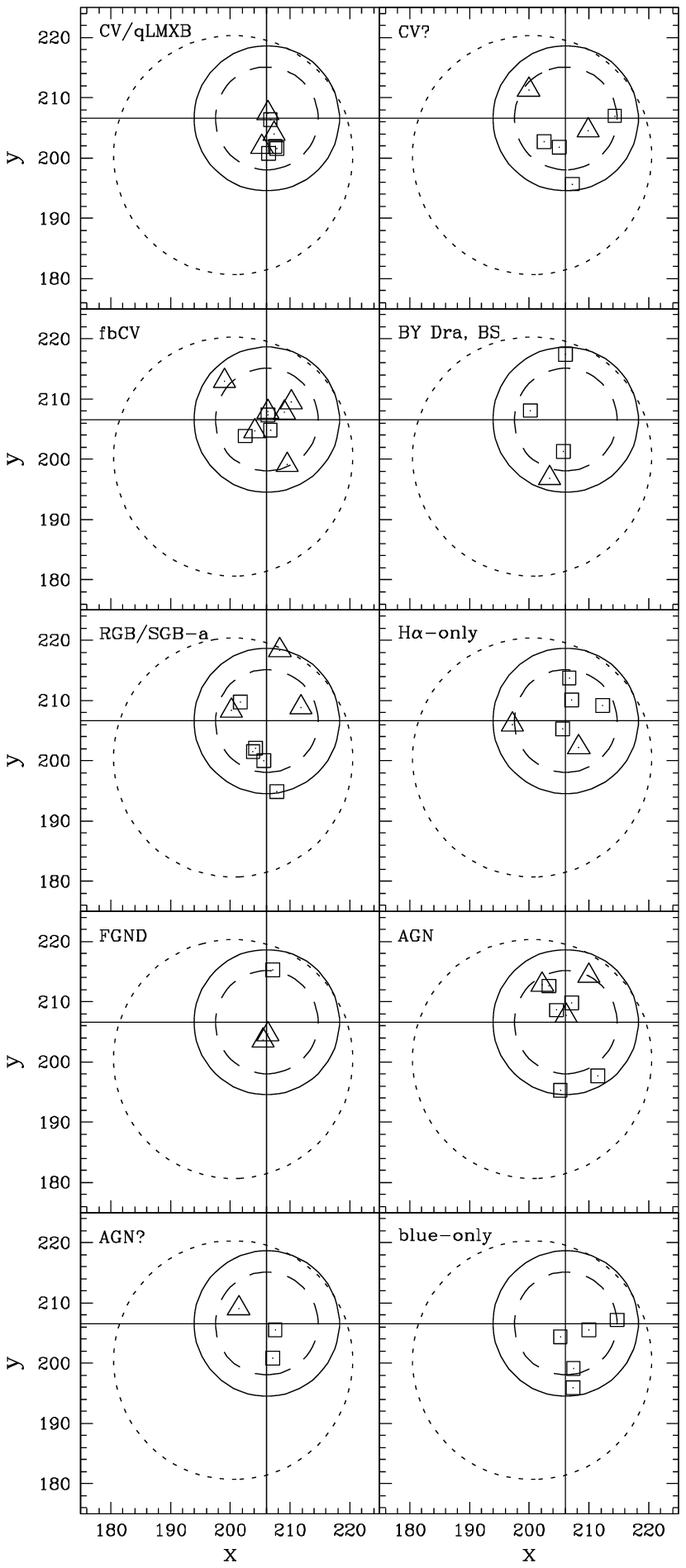}
}
\figcaption{Location within X-ray error circles of optical counterparts 
listed in Table 1.  Dotted line shows initial 1\asec-radius error circle, solid line
shows 0.6\asec-radius error circle adopted for detailed search.  Dashed line 
encompasses half of the area of the 0\secspt 6 error circle (see \S 2.2 for details).
Symbols as in Fig.\ 2.}
\end{figure}

As an additional test of the validity of potential counterparts we
examined where they landed in the X-ray error circles.  Fig.\ 4 shows
the location of each of the 59 possible optical counterparts listed in
Tables 1 and 2 and shown in Fig.\ 3, dividing the objects into 10
categories (see column 8 of Table 2).  In each of the 10 panels we
show the original 1\asec\ X-ray error circle (dotted line), the final
0\secspt 6 circle (solid line with cross hair), and a smaller circle
of radius 0\secspt 424 encompassing half of the area of the 0\secspt 6
error circle (dashed line).  We refer to these results in \S 3 below,
as we examine each category of optical counterpart in turn.

% SECTION 3

\section{Identification of Optical Counterparts}

% SECTION 3.1

\subsection{Blue, \ha-bright stars:  cataclysmic variables and a qLMXB} 

We first consider sources for which we have identified an optical
counterpart that is both blue and \ha-bright.  This combination of
signatures is strongly indicative of a compact accreting binary, with
the \ha\ excess attributable to an emission line from an accretion
disk and the blue color to the disk and/or the white dwarf.  Optical
counterparts for 13 of the X-ray sources have this signature.  The
first of these, shown in panel 1 of Fig.\ 3, is the qLMXB reported by
Haggard et al.\ (2004).  This object was first identified as a qLMXB
on the basis of its X-ray spectrum (Rutledge et al.\ 2002) and is the
only X-ray source in \wcen\ with both the soft spectrum and the
luminosity characteristic of qLMXBs (HCD09).

The next three panels in Fig.\ 3 show three objects that were first
identified in our WFPC2 study of \wcen\ (Carson et al.\ 2000).  Panels
2 and 3 show the optical counterparts of ROSAT sources ``XA'' and
``XB,'' respectively, both of which Carson et al.\ identified as
probable CVs on the basis of significant \ha\ and UV excesses.  Both
are clearly \ha-bright in the ACS data and 13c (=XA) is also very
blue.  Interestingly, however, 13a (=XB) is only very slightly blue
relative to the main sequence in the new data.  This suggests a
relatively weak disk in the system and/or significant variability.  A
third star identified by Carson et al.\ as a possible counterpart for
ROSAT source ``XC'' (\Chandra\ source 12a) is shown in panel 4 (also
see Fig.\ 2).  In the 1997 WFPC2 data this star was \ha-bright but
showed no UV excess (Carson et al.\ 2000).  Here it appears distinctly
blue as well as being \ha-bright.  We conclude that it is likely to be
a CV; variability could account for the apparent change in color
between 1997 and 2002.

In panels $5-7$ of Fig.\ 3 we show, in order of \r\ magnitude, three
new CV candidates that are both blue and \ha-bright.  They are the
counterparts of X-ray sources 54h, 41d, and 24c.  The optical
counterpart to 54h is easily confirmed visually as being blue, and all
the individual \hr\ measurements (from single \ha\ and \r\ frames)
consistently show it is \ha-bright.  The counterpart to 41d is easily
visually confirmed as being \ha-bright, and the individual \br\
measurements consistently show it is blue.  The counterpart to 24c is
visually confirmed as being both blue and \ha-bright.  These three
stars are listed along with the optical counterparts of X-ray sources
13c, 13a, and 12a as ``CVs'' in Tables 1 and 2.

Six additional CV candidates are shown in panels $8-13$, in order of
increasing magnitude, from \r $= 22.0-25.2$.  Each of these stars is
visually confirmed as being either blue or \ha-bright.  Somewhat lower
confidence in either their blue color or \ha\ excess (see Table 2,
columns 5 and 6) leads us to give them the more tentative designation
``CV?'' in Tables 1 and 2.

The top left panel of Fig.\ 4 shows the positions within their
respective error circles of the six most secure CV counterparts along
with the qLMXB (44e).  These stars are tightly clustered near the
center of the error circles and well inside even the inner dashed
circle, leaving little doubt that they are the counterparts of the
X-ray sources.  The top right panel shows the six tentative CV
counterparts.  While not as tightly clustered in the centers of the
error circles, five of the six land inside the inner, dashed circle
(whose area was chosen to be one half that of the 0.6\asec\ circle),
the probability of which is 6\x (0.5)$^6$ = 9.4\% if the stars were
unrelated to the X-ray sources.  Larger positional uncertainties are
also to be expected for a population of objects that are fainter on
average in both X-ray and optical light.  This further supports our
conclusion that these are the probable counterparts of the X-ray
sources.

\begin{figure}
\vspace{-3mm}
%\plotone{f5.eps} 
\centerline{
\includegraphics[scale=0.55]{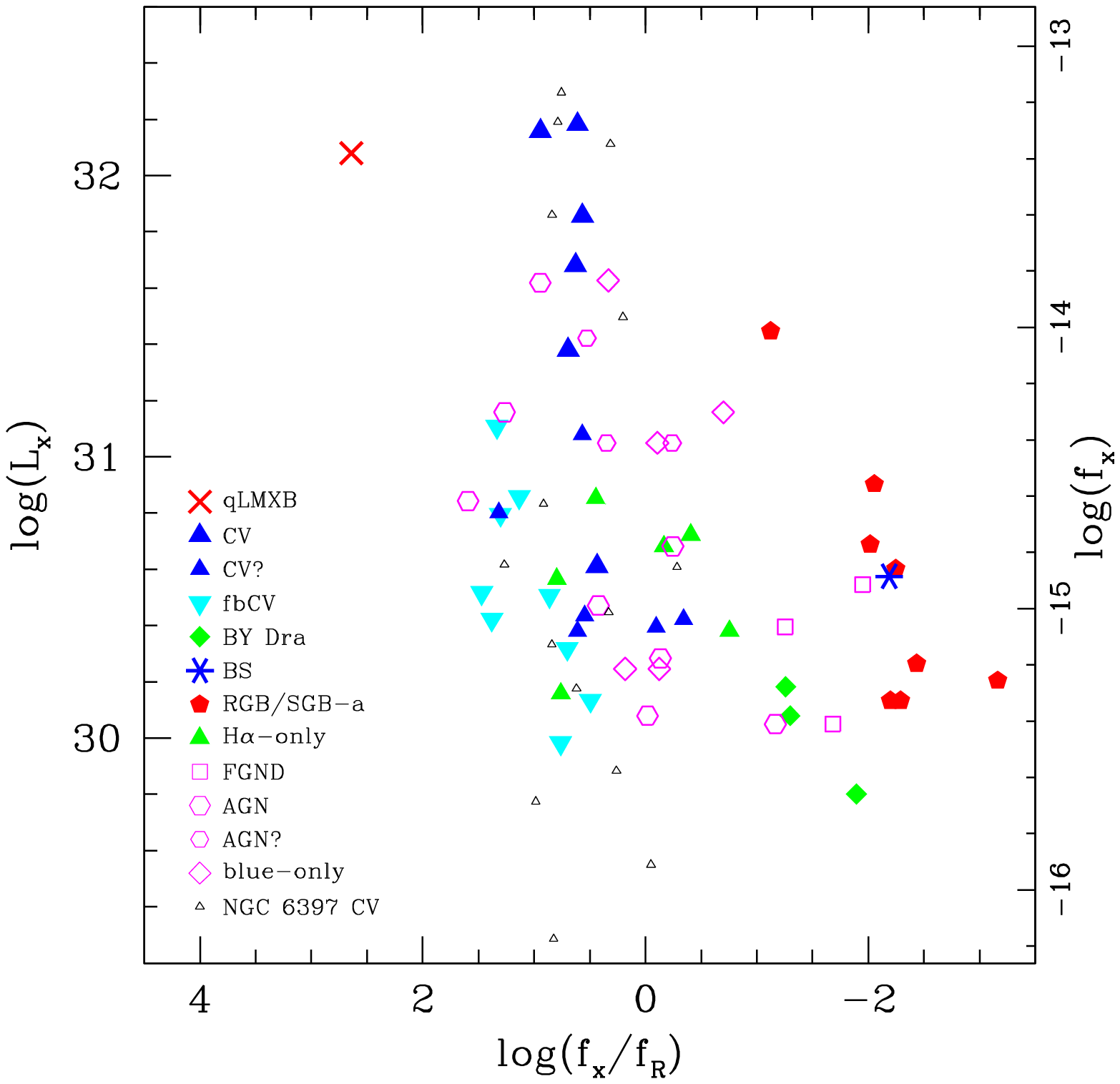}
}
\vspace{-1.5cm}
\figcaption{X-ray luminosity in \ergs\ and flux in
  \ergscmsq\ vs.\ X-ray--to--optical flux ratio for optical counterparts
  identified using \HST\ data.  X-ray fluxes and luminosities are in a
  $0.5-2.5$ keV band, adopted from HCD09, but revised downward by a
  factor of 1.25 (see \S 4.1).  Large solid symbols represent
  candidates likely to be members of \wcen\ while large open symbols
  represent non-members (and probable non-members).  For comparison,
  small black open triangles are CV candidates identified by Cohn et 
  al.\ (2010) in NGC~6397; only the \lx\ scale applies for these points.}
%\vspace{-5mm}
\end{figure}

Further insight into the nature of the potential optical counterparts
can be gained from their X-ray--to--optical flux ratios and their
luminosities.  In Fig.\ 5 we plot the X-ray luminosity of each object
against the ratio of its X-ray to \r-band flux.  X-ray fluxes are in
the $0.5-2.5$ keV band and assume a 1 keV thermal brehmsstrahlung
spectrum (HCD09); luminosities were computed assuming a distance to
the cluster of 4.9 kpc.  Large blue triangles denote the six CV
candidates with clear-cut blue colors and \ha\ emission.  Smaller blue
triangles denote the tentative CV candidates.  Four of these have
flux ratios very similar to those of the other CVs.  The other two
have somewhat lower flux ratios, but still resemble the CVs in this
diagram more than other types of systems (e.g. active binaries, which
have much lower X-ray--to--optical flux ratios; see below).  We
conclude that these six optical counterparts are also likely to be
CVs.

% SECTION 3.2

\subsection{Faint blue stars:  more cataclysmic variables?}

Very faint blue stars were found in the error circles of nine of the
\Chandra\ sources.  They have magnitudes in the range \r $= 24.5-26.3$
and are shown in Fig.\ 3, panels $14-22$.  All but one (54e) are
confirmed visually as being blue by blinking \r\ vs.\ \b\ images.
None of these stars are detected in \ha\ with ALLFRAME---thus they do
not appear in the \hr\ diagrams.  These stars lie in the region of the
CMDs generally occupied by white dwarfs; in the absence of X-ray
emission one might simply assume that is what they are.  Thus we need
to consider whether this many white dwarfs could have landed by chance
in the \Chandra\ error circles.  The 109 0\secspt 6 radius error
circles occupy about 0.03\% of the 10\amin \x\ 10\amin\ field.
Monelli et al.\ (2005) used these data to study the WD sequence and
found 2212 WDs in 1/3 of the total area covered.  One would therefore
expect \about 3 \x\ 2212 \x\ 0.0003 $\simeq$ 2 WDs to fall within the
\Chandra\ error circles by chance.  Considering that all but one land
in the inner half of the error circle (see Fig.\ 4, ``fbCV'' panel),
this number reduces to \about\ 1.  The probability that this would
happen if the stars were unassociated with the X-ray sources is 1.8\%.
We conclude that these stars are the likely sources of the X-rays and
not simply WDs that have landed by chance in the X-ray error
circles.\footnote{The fact that these stars lie within the inner half
of their respective error circles also confirms that the adopted error
circle radius of 0\secspt 6 is sufficiently generous even for faint
X-ray sources.  The new distortion correction confirms this
conclusion, with 8 of the 9 lying less than 0\secspt 35 arcseconds
from the newly determined mosaic positions.}

These nine stars are exceedingly faint, with absolute magnitudes in
the range \Mr\ $\simeq$ $10.7-12.5$ at the distance of \wcen.  Most
are seen in \b\ only because they are so blue; main sequence stars of
comparable \r\ magnitude are below the detection limit.  In view of
this, the lack of \ha\ detection does not necessarily imply that they
are not \ha-bright; they may simply be too faint to be detected, even
in the presence of an \ha\ emission line.  Indeed, only one of the
moderately bright CV candidates (44d), is fainter than even the
brightest of these nine stars, and its \ha\ status is uncertain.  The
faintest CV candidate for which excess \ha\ emission is clear cut
(24c) is 1 magnitude brighter than the brightest of these stars and
$2-3$ magnitudes brighter than most of them.  As discussed in \S 4
below, we suggest that these stars are likely to be very faint
cataclysmic variables near the period limit, similar to those
identified in the Sloan Digital Sky Survey (SDSS; G\"{a}nsicke et al.\
2009) and in NGC~6397 (Cohn et al.\ 2010).  Their positions in the
X-ray luminosity vs.\ \fxfr\ diagram (inverted light blue triangles in
Fig.\ 5) are also consistent with this interpretation.  The median
value of the ratio of soft to hard X-ray counts reported by Haggard et
al.\ (2009) for these sources (see their Table 1) is 1.2.  This is
consistent with their being CVs (cf.\ Fig.\ 6 of Haggard et al.\
2009), and argues against the possibility that they could instead be
MSPs with low-mass white dwarf companions (which can appear in a
similar part of an optical CMD---see Edmonds et al.\ 2001); MSPs
typically have much softer X-ray colors (see, e.g., Fig.\ 9 of Heinke
et al.\ 2005).  We refer to these stars hereafter as ``faint blue
CVs'' (fbCVs).

% SECTION 3.3

\subsection{BY~Draconis stars and a possible blue straggler}

Narrow-band \ha\ imaging also enables us to search for binaries in the
form of BY~Draconis stars.  These systems show \ha\ in emission due to
elevated levels of coronal activity resulting from rapid spin rates
typically as a result of tidal synchronization with a companion star
(Dempsey et al.\ 1997 and references therein).  Such stars have been
found in a number of globular clusters to date (e.g., Taylor et al.\
2001, Albrow et al.\ 2001, Pooley et al.\ 2002, Huang et al.\ 2010,
Cohn et al.\ 2010).

Considering the typical emission line strengths of such systems
(EW(\ha) \simlt 5 Angstrom, e.g. Chevalier \&\ Ilovaisky 1997), and
the \about 80 Angstrom width of the ACS/WFC \ha\ filter, such stars
will have \ha\ excesses (relative to normal main-sequence stars)
\simlt 0.06 magnitudes in the present study.  This is sufficiently
small that we expect only to be able to detect the subset with the
strongest lines.  To limit the number of false positives, we required
that a star show both an \ha\ signature and also lie redward of the
main-sequence ridge line in the \br\ CMD.  The latter requirement
excludes binaries whose mass ratios are much less than unity.  The
three possible BY~Dra stars found in this way are shown in Fig.\ 3,
panels $23-25$, and are designated ``BYDra'' in Table 2.  The \fxfr\
ratios for these stars are much lower than for accretion-driven
systems, on the order of 10$^{-2}$ (see green diamonds in Fig.\ 5),
consistent with their tentative identification as stars with active
coronae.  Their X-ray luminosities (\about $10^{30}$ \ergs) are also
consistent with their being coronal sources (cf. Dempsey et al.\
1997).

A further test of the viability of these three counterparts can be
made by examining the ratio of their X-ray to bolometric luminosities,
a quantity that has been shown to depend on stellar rotation and to
saturate at a maximum value of log(\lx/L$_{\rm bol}$) \about $-3$
(Stauffer et al.\ 1994, G\"{u}del 2004, and references therein).
Using bolometric luminosities derived from the stellar models of Bedin
et al.\ (2005) appropriate for the dominant metallicity group in
\wcen, together with the \lx\ values given in Table~2, we find that
the proposed optical counterparts for 24e, 43e, and 14a would have
log(\lx/L$_{\rm bol}$) $= -3.1, -2.5$, and $-2.4$, respectively.  That
two of the three have ratios that are a factor of \about 3 higher than
the known limit suggests that at least some of these stars may not be
the counterparts of the X-ray sources.\footnote{Correcting for the
differing bandpasses in which \lx\ values are reported by, e.g.,
Stauffer et al.\ (1994) vs.\ Haggard et al.\ (2009) increases the
ratios we obtain by an additional factor of \about 1.5, making the
discrepancy worse.}  Given the many uncertainties that go into
computing these ratios (e.g., the X-ray sources in question are very
faint), we report the counterparts nonetheless, as they remain the
best prospects we have found for BY~Dra stars in \wcen.

\begin{figure}
\vspace{-3mm}
%\plotone{f6.eps} 
\centerline{
\includegraphics[scale=0.55]{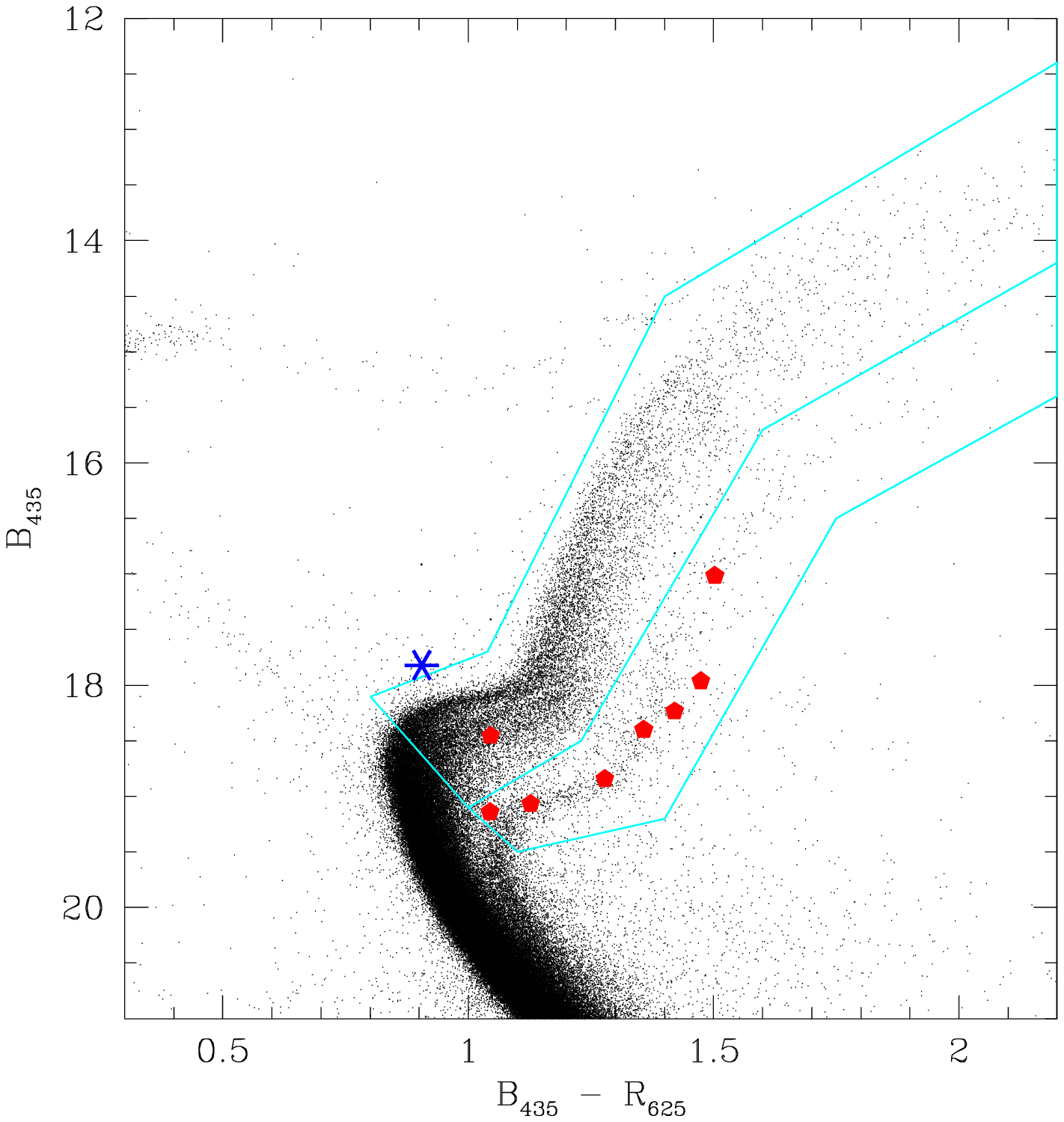}
}
\vspace{-1.6cm}
\figcaption{Close-up of the turnoff, subgiant, and giant regions of
  \wcen, from the photometry of Anderson \&\ van der Marel (2010).  To
  best delineate the different RGB and SGB sequences, we plot only
  about half of the stars.  Seven of the optical counterparts we have
  identified lie on or near the anomalous RGB or SGB, which is
  significantly redder than the other RGB and SGB sequences in the
  cluster.  For simplicity, we include an eighth star (counterpart of
  13b) in the classification although it lies along one of the other
  subgiant sequences.  The blue asterisk marks a star identified as a
  possible turnoff binary or blue straggler (see \S 3.3).  Regions
  outlined in cyan are used to determine the significance of having
  found 7 X-ray-bright stars apparently associated with the RGB/SGB-a
  sequence (see \S 3.4).}
\end{figure}

Another candidate optical counterpart (22d; see panel 26) appears
above the turnoff, to the blue side of the subgiant branch.  A
close-up of this region is shown in Fig.\ 6, based on photometry of
1.2 million stars in the cluster from Anderson \&\ van der Marel
(2010); for clarity, only half the stars are shown in this plot.  The
star lies directly above the turnoff associated with the dominant
population of stars in the cluster.  Its \fxfr\ flux ratio and X-ray
luminosity (see blue asterisk in Fig.\ 5), and log(\lx/L$_{\rm bol}$)
$= -3.5$, are all within observed values for coronal sources.  This
suggests that it could be a rapidly-rotating blue straggler, a blue
straggler with an active main-sequence star companion (see, e.g.,
Knigge et al.\ 2006), or perhaps a BY-Dra-type system that contains
two turnoff stars.  Notably, however, the star has a rather large \ha\
excess (0.25 mag; see Fig.\ 3 panel 26), corresponding to an emission
line strength of EW(\ha) \about 20 Angstroms.  This is unusually
strong for coronal sources, for which EW(\ha) is typically an order of
magnitude or more lower (cf.\ Young et al.\ 1989).  Alternatively, it
could be a CV in which the white dwarf is accreting from a subgiant.
In this case the high optical luminosity of the subgiant might account
for the rather low \fxfr\ ratio.

% SECTION 3.4

\subsection{Stars on or near the anomalous subgiant and giant branch}

Eight of the candidates we have identified lie redward of and/or below
the giant and subgiant branches formed by the dominant population in
the cluster (the most metal-poor population, with [Fe/H] $= -1.7$;
``group A'' in the nomenclature of Villanova et al.\ 2007).  They have
magnitudes in the range \r $= 15.5-18.1$ and colors in the range \br\
$= 1.1-1.5$ (see Fig.\ 3, panels $27-34$).  To assess the possible
connection between these stars and the complex of giant and subgiant
branches in \wcen, in Figure~6 we plot a CMD for the full ACS/WFC
field from the photometry of Anderson \&\ van der Marel (2010).  Here
the multiple giant and subgiant branches are well populated and
readily identified.  The eight stars of interest all appear in the
catalog created by Anderson \&\ van der Marel (2010) and are marked
with red pentagons.  Seven lie on or very close to the so-called
anomalous red giant and subgiant branches in the cluster (RGB-a and
SGB-a; Pancino et al.\ 2000, Ferraro et al.\ 2004).  Given their
locations in the CMD, we designate these stars ``RGB/SGB-a'' in Table
2.  However, as discussed below (see \S 4.2), in the absence of
metallicity measurements, it is unknown whether these stars are actual
members of this metal-rich ``anomalous'' population in \wcen, or if
they lie in this part of the CMD for other reasons.  For simplicity we
also include in the ``RGB/SGB-a'' category the eighth star, a possible
counterpart of 13b (Fig.\ 3, panel 30), though it is clearly not a
member of the anomalous population.  It lies along the subgiant branch
associated with the ``group C'' population identified by Villanova et
al.\ 2007, and is \about 0.3-0.4 magnitudes fainter in \b\ than the
(``group A'') giant and subgiant branches (Fig.\ 6).

Before addressing the potential significance of these stars, we must
determine whether their apparent association with X-ray sources could
be the result of chance coincidence.  Adopting the lower outlined
region shown in Fig.\ 6 that roughly encompasses the SGB-a and RGB-a
stars, we find that there are a total of \about 3000 such stars in the
ACS/WFC field of view.  Given that all of the 0\secspt 6 error circles
combined cover only \about 1/3000 of the total area of the mosaic, we
would expect \about 1 of these stars to have landed by chance in the
X-ray error circles.  That 7 of them appear in X-ray error circles
strongly supports the view that they are the sources of the X-rays
(chance probability $\simeq$ 2\x $10^{-5}$).

Two additional arguments favor these stars being the sources of the
X-ray emission.  First, three show signs of \ha\ in emission, which is
strongly associated with enhanced X-ray emission.  Second, the X-ray
luminosities and \fxfr\ ratios for these stars are typical of stars
with active coronae (see red pentagons in Fig.\ 5).

Having demonstrated that they are the likely sources of the observed
X-ray emission, we must also ask whether these stars are associated
with the cluster.  Proper motions can be used to determine membership,
but require measurements in two well-separated epochs of imaging,
which is beyond the scope of the present paper.\footnote{The only one
of these stars (13b) that appears in the proper-motion catalog of
Anderson \&\ van der Marel (2010) has motions consistent with
membership.}  In the absence of proper motions, we consider instead
what other explanation there could be for a group of stars in this
region of the CMD.  As they are redder than typical stars in \wcen,
one possibility is that they could be foreground stars.  However, it
is highly improbable that such a large number of foreground stars
would be confined to a small region of color-magnitude space.  At
fainter magnitudes, where foreground stars would be expected to be
more prevalent given the larger volume encompassed, we find only two
foreground stars (see \S 3.6).  The stars in question also have \br\
and \hr\ colors quite different than those of the faint foreground
stars (see \S 3.6).  We conclude that the stars that we have
identified in the vicinity of the RGB/SGB-a are likely to be
associated with \wcen.

It is also of interest to determine whether the frequency of X-ray
bright stars among RGB/SGB-a stars is larger than among other giants
and subgiants in the cluster.  To this end, we counted the number of
stars in the upper outlined region shown in Fig.\ 6, that encompasses
groups A, B and C as defined by Villanova et al.\ 2007 (i.e., all
giants and subgiants that are not part of the anomalous branches).  A
total of \about 30,000 stars lie in this portion of the CMD.  Of
these, we would expect about 1 in 3000, or \about 10, to have landed
in the X-ray error circles by chance.  Searching through the CMDs we
constructed for all the X-ray source error circles, we find 14
``normal'' giants in total, consistent with them all being chance
alignments.  Thus, despite the fact that there are \about 10 times
more stars on the upper subgiant and giant branches, only twice as
many land in X-ray error circles than do the SGB-a and RGB-a stars.
This implies that X-ray sources are at least 5 times overabundant
among the SGB-a and RGB-a stars relative to the normal giant and
subgiant branch stars.  Taking account of the fact that only \about 4
of the normal RGB/SGB stars found in error circles are likely to be
true X-ray source counterparts (vs.\ \about 6 of the anomalous RGB/SGB
stars), the implied overabundance rises to a factor of \about 15.  The
significance of these stars and their possible association with stars
found in similar regions of the CMD in other clusters (e.g., so-called
``red stragglers'' and ``sub-subgiants'') will be discussed below (see
\S 4.2).

% SECTION 3.5

\subsection{Stars with \ha\ excess only: CVs or BY Dra binaries?}

Six stars were identified that show signs of \ha\ in emission but land
on the main sequence in the \br\ CMD---or, in one case, are not
detected in the blue filter (see panels 35-40 in Fig.\ 3).  While the
emission is quite weak in some cases, the fact that five of these six
possible counterparts lie in the inner half of the error circles (see
Fig.\ 4) suggests that most are likely to be genuine counterparts of
the X-ray sources.  In Tables 1 and 2 we designate these stars as
``\ha-only.''

Objects with this \ha\ signature could be CVs with weak disks, such as
have been seen in NGC~6397 (Cool et al.\ 1998, Cohn et al.\ 2010).  In
\wcen\ itself, the optical counterpart of source 13a, which was
previously determined to be a CV (Carson et al.\ 2000), is only very
slighly blue in the present data, and the counterpart of 12a, seen to
be quite blue here, was formerly seen to be on the main sequence
(Carson et al.\ 2000).  Some of these stars could alternatively be BY
Dra-type main-sequence binaries, in which the secondary star is too
faint to noticelably alter the color of the combined system.

A potentially distinguishing feature between X-ray sources powered by
accretion and those resulting from active coronae is the ratio of
X-ray--to--optical flux.  In Fig.\ 5 we see that three of the \ha-only
stars (green triangles) lie solidly in the region occupied by CVs.
Two others are near the edge of the CV region, while the third lies
between the CV region and the region occupied by coronal sources.  We
conclude that the perhaps 4 or 5 of these sources are likely to be
CVs, while one or more could instead be BY Dra-type binaries.

% SECTION 3.6

\subsection{Foreground stars}

Three of the optical counterparts have magnitudes and colors that
place them $\simgreat$ 1 magnitude redward of the cluster main
sequence or turnoff (see Fig.\ 3, panels 41-43).  Their \br\ colors
are in the range $2.1-2.5$, suggestive of late K or early M stars.
All three also have significant \ha\ excesses (\about 0.15-0.5
magnitudes) and X-ray--to--optical flux ratios that are typical of
coronal sources (see magenta squares in Fig.\ 5).  We conclude that
these are most likely foreground dwarfs with active coronae (e.g., dMe
or dKe stars), which are very common in the field (Riaz, Gizis, \&\
Harvin 2006).  Two such stars have already been identified in the
foreground of \wcen\ (Cool et al.\ 1995a).  We note that the brightest
of the three (43f) has an \r\ magnitude of 17.7, in the same range as
the RGB/SGB-a stars discussed in \S 3.4.  However, its very red \br\
color and prominent \ha\ emission make it appear more like the two
other fainter foreground stars and we therefore include it in this
class.

% SECTION 3.7

\subsection{Active galactic nuclei}

Eight of the X-ray source error circles contained visibly extended
objects.  Given the large number of AGN we expect in the background of
this \wcen\ mosaic (HCD09), it seems likely that most or all of these
are the optical counterparts of the X-ray sources.  In most cases no
other compelling optical counterpart was found in the error circle.
We note that AGN can be useful in determining the absolute motions of
globular clusters.

The locations of the AGN within the CMDs are shown in Fig.\ 3, panels
$44-51$; magnitudes and colors are necessarily approximate given that
we derived them from fitting the PSF to an extended source.  Half of
the extended sources have a distinctive blue, \ha-faint signature
(52b, 54d, 31b, 62b); one of these (52b) was double-peaked in the
image and appears as two triangles in the CMDs.  Seven of the eight
have relatively high X-ray--to--optical flux ratios (see large magenta
hexagons in Fig.\ 5), consistent with their identification as
accretion-powered X-ray sources.

Three additional objects were found in the X-ray error circles that
appeared to be point-like, but had the same blue, \ha-faint signature
found for half of the extended sources.  These objects are shown in
panels 52-54 of Fig.\ 3.  Given that this combination of blue color
and \ha\ deficit was uniquely associated with several clearly extended
objects, we tentatively identify these objects as AGN, and designate
them ``AGN?'' in Tables 1 and 2.  Their X-ray--to--optical flux ratios
are also consistent with this designation (see small magenta hexagons
in Fig.\ 5).  It is likely that the crowding and high background
levels in the images could mask any extended emission that might be
present.

% SECTION 3.8

\subsection{Other blue stars}

Blue stars with \r $= 19.3-23.8$ were found in the error circles of
five \Chandra\ sources (see Fig.\ 3, panels $55-59$).  In contrast to
the very faint blue stars discussed above, these stars are bright
enough to be readily detected in \ha.  All fall on the main sequence
in \hr\ and thus show no sign of \ha\ in emission.  In Tables 1 and 2
we designate these stars as ``blue-only.''

The two faintest of these (33c and 24a) are in a region of the diagram
that is populated by numerous background stars.  The most probable
explanation for these is that they are background stars that have
landed by chance in the X-ray error circles.

The three brighter stars (33h, 44a, and 51a), by contrast, are in a
region of the CMD that is quite sparsely populated, and are thus much
less likely to be chance alignments.  Two have \br\ colors
significantly bluer than the turnoff, making it improbable that they
are background stars.  One possibility is that they are CVs with weak
emission lines.  At the magnitudes of these stars, EW(\ha) would need
to be \simgt 5-10 Angstroms to produce a detectable \ha\ excess.
Emission lines this weak are not unknown among CVs.  Alternatively,
one or more of these objects could be AGN.  Either interpretation
would be consistent with their \fxfr\ values, which are similar to
those of CVs or AGN (see open magenta diamonds in Fig.\ 5).

% SECTION 4

\section{Discussion}

The ACS/WFC mosaic encompasses 109 of the 180 known X-ray sources in
and toward \wcen.  Using \b, \r, and \ha\ imaging we have identified
promising optical counterparts for more than half of these sources.
The optical counterparts divide into several categories.  Probable
cluster members include cataclysmic variables, a quiescent low-mass
X-ray binary, possible BY Draconis stars, and a possible blue
straggler.  They also include a new class of sources that appear to be
associated with the cluster's anomalous RGB and SGB populations.
Finally, we identify three probable foreground stars and at least
eight AGN behind the cluster.

% SECTION 4.1

\begin{figure}
%\plotone{f7.eps} 
\centerline{
\includegraphics[scale=0.45]{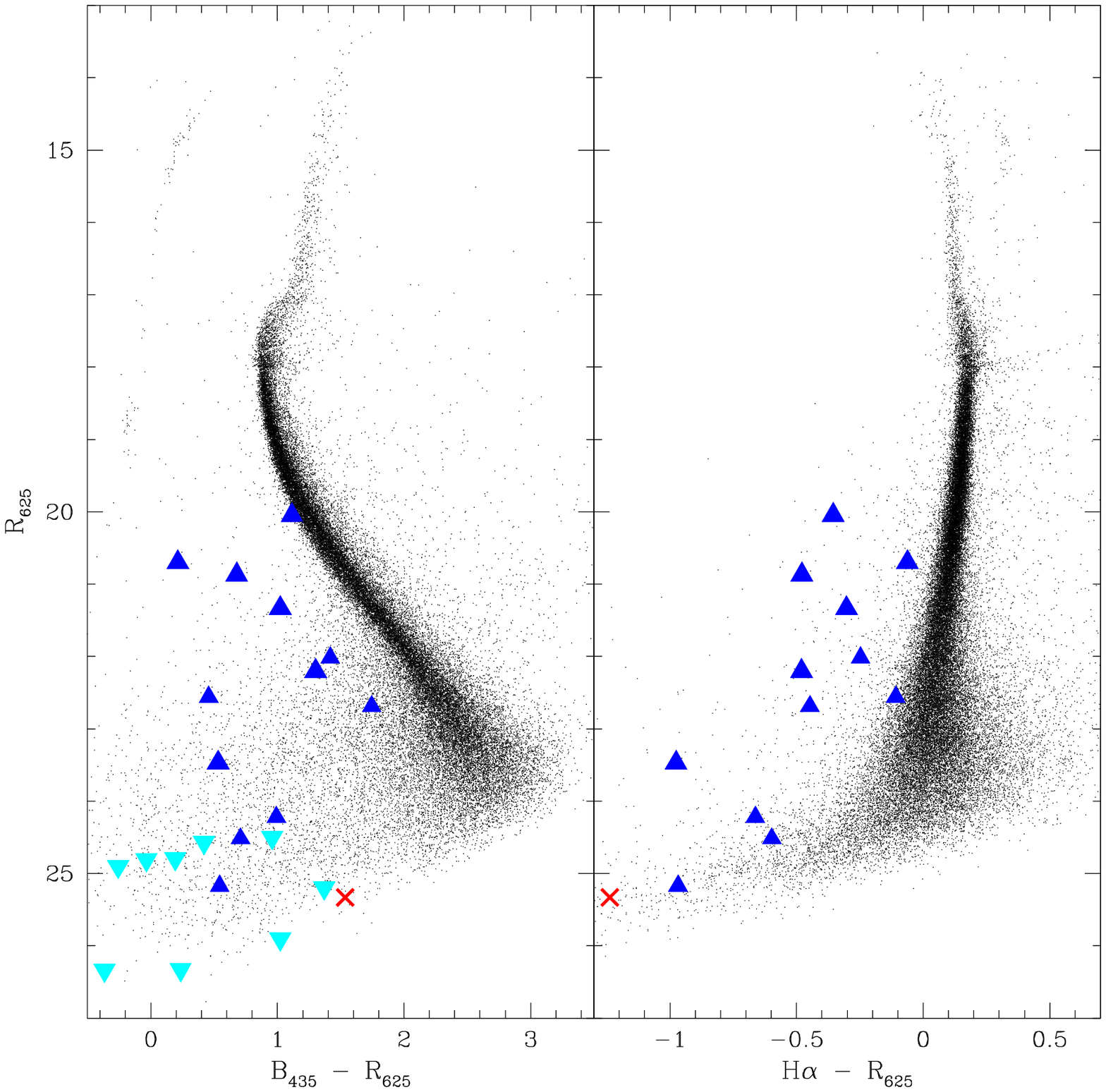}
}
\figcaption{Color--magnitude diagrams showing 12 CV
  candidates that are both blue and \ha-bright, together with 9 that
  are very faint and blue but undetected in \ha.  The qLMXB is also
  shown.  Symbols as in Fig.\ 5.  Six more tentative ``\ha-only'' CV
  candidates, some of which could be BY~Dra stars (see \S 3.5), are
  not shown.}
\vspace{2mm}  
\end{figure}

\subsection{The CV population in \wcen}

The single largest class of optically identified X-ray sources in
\wcen\ is cataclysmic variables (see Fig.\ 7).  We find 27 candidates,
only three of which were previously known (Carson et al.\ 2000).  This
is among the largest number of CV candidates yet identified in a
globular cluster, comparable to the population found in 47~Tuc
(Edmonds et al.\ 2003a,b; Heinke et al.\ 2005).  The candidates span
more than 6 magnitudes in apparent brightness, with \r\ $= 20.0-26.3$.
Adopting a distance modulus of (m$-$M)$_{625}$ = 13.8 (Haggard et al.\
2004), this corresponds to absolute magnitudes in the range \Mr
$=6.3-12.6$.  There is a hint of bimodality in the distribution (blue
and cyan symbols in Fig.\ 5), possibly indicative of a period gap,
with just one CV in the range \r\ $=23-24$ (\Mr\ $\simeq 9.2-10.2$),
and more than half of the candidates being fainter than \r\ $=24$.

The absolute magnitudes of the faint CV candidates are similar to
those of the faintest CVs discovered in the SDSS (Szkody et al.\
2011), which have been shown to have orbital periods of $80-86$ min,
at or near the theoretical limit for CVs (G\"{a}nsicke et al.\ 2009).
That discovery resolved a long-standing controversy concerning the
theory of late-stage evolution of CVs, which predicts a pile up of
old, faint systems near the period minimum.  The subset of these
faint, short-period SDSS CVs for which distances have been determined
have absolute magnitudes in the range M$_g = 10.5-13.1$; given their
$g-r$ colors (G\"{a}nsicke et al.\ 2009), their $V$-band magnitudes
should be similar.  This is to be compared to \Mr\ $= 10.4-12.6$ for
the \wcen\ CVs, for which we expect colors in the range $V-R$ \about\
$0.0-0.5$, given their observed \br\ colors.  The similarity in
absolute magnitudes of these systems suggests that we have identified
a population of short-period, low-accretion-rate CVs in \wcen.

\wcen\ is the second globular cluster in which such a faint population
of CVs has been found.  A similar group was identified in NGC~6397 by
Cohn et al.\ (2010), with median \Mr\ $=$ 11.2.  Notably, the faint
CVs in NGC~6397 lie on or close to the WD sequence in a \br\ vs.\ \r\
CMD (see Fig.\ 3 of Cohn et al.), suggesting that the optical light is
dominated by the white dwarf.  This is similar to what we observe in
\wcen, though in \wcen\ the color spread is larger, due at least in
part to measurement uncertainties, as these stars are close to the
magnitude limit of the data.  In contrast to isolated WDs, the
luminosity of the WD in a CV is not a measure of its age, but is
instead a measure of the time-averaged accretion rate onto the WD
(Townsley \&\ Bildsten 2002).  If we assume that the optical flux of
the faint CV candidates in \wcen\ is dominated by the WDs, the
corresponding accretion rates predicted by Townsley \&\ Bildsten
(2002) are $10^{-11}-10^{-9}$ \msun/yr.  Such accretion rates are
consistent with periods below the period gap (Patterson 1984; Townsley
\&\ G\"{a}nsicke 2009; Knigge, Baraffe, \&\ Patterson 2011).

The lack of observed \ha\ emission among many of the faintest CVs in
\wcen\ (the 9 ``faint blue CVs''---see Tables 1 and 2) can also be
understood if their spectra are similar to those of the faint CVs in
the SDSS.  A key spectral signature of SDSS CVs near the period
minimum is the presence of broad hydrogen absorption lines superposed
on the emission lines that are prototypical of CVs (G\"{a}nsicke et
al.\ 2009).  This dominance of the WD in the spectrum not only makes
the systems very blue (i.e., they lie on or near the WD sequence in a
\br\ vs. \r\ CMD), but also reduces the excess \ha\ flux relative to
the continuum.  This alters the \hr\ colors measured from the imaging
data in such a way that they appear to have little or no \ha\ excess.
In NGC~6397, the effect is particularly noticeable for the five CV
candidates that lie on or slightly blueward of the WD sequence.  Those
five candidates show no \ha\ excess relative to the main sequence.
However, they are still clearly \ha-bright by comparison to the WD
sequence, which appears as a group of faint stars to the \ha-faint
side of the main sequence in the \hr\ vs.\ \r\ diagram (also see
Strickler et al.\ [2009] for a discussion).  Given the relative
shallowness of the \wcen\ \ha\ data as compared to those obtained for
NGC~6397, the lack of observed \ha\ emission among the ``fbCVs''
appears to be compatible with their being similar to the faint CVs in
NGC~6397.

Further insight into the nature of the CVs in \wcen\ can be gained by
examining their X-ray--to--optical flux ratios.  This
distance-independent quantity can be directly compared to values for
CVs in other clusters and in the field.  It is of particular interest
to compare these ratios to those for CVs in NGC~6397, since both are
X-ray selected samples collected using the same methods.  To compute
flux ratios for the CVs in \wcen, we adopt the unabsorbed \fx\ values
reported by HCD09, revised downward\footnote{Here we use
http://asc.harvard.edu/toolkit/pimms.jsp to convert count rate
to flux; fluxes reported by HCD09 used WEBPIMMS, and did not account
for the higher sensitivity of the instrument in early years of
operation (\wcen\ X-ray data were acquired in 2000).} by a factor of
1.25, and \fr\ = 10$^{0.4R-5.89}$.  The latter makes use of the
average flux per unit wavelength of Vega in the F625W
band,\footnote{http://www.stsci.edu/hst/acs/analysis/zeropoints}
the $PHOTBW$ width of the
filter,\footnote{http://www.stsci.edu/hst/acs/analysis/bandwidths}
and includes a correction for an estimated 0.29 magnitudes of
extinction toward \wcen\ (Haggard et al.\ 2004).  The resulting flux
ratios for the CVs in \wcen\ are in the range \fxfr\ $\simeq 0.2-30$,
with a median value \fxfr\ $=$ 4.2 (see Fig.\ 5).

To make a comparison to NGC~6397, we need to account for the different
assumptions used to derive X-ray and optical fluxes for CVs in the two
clusters.  This requires multiplying the X-ray fluxes reported by
Bogdanov et al.\ (2010)\footnote{Beginning with the $0.5-6.0$ keV
counts reported by Bogdanov et al. (2010) (in an effective exposure
time of 237 ksec---S.\ Bogdanov, private communication), and assuming
the same spectrum and band as for the \wcen\ CVs, we find that 1
count/sec is equivalent to an unabsorbed flux of 5.81 \x\ $10^{-12}$
\ergscmsq\ if we use the same \NH\ adopted by Bogdanov et al.  This is
a factor of 1.17 higher than the conversion adopted in that paper for
a different set of assumptions.} by a factor of 1.17 and dividing
optical fluxes reported by Cohn et al.\ (2010) \footnote{Optical
fluxes reported by Cohn et al.\ (2010) assume \fr\ = 10$^{0.4R-6}$,
which results in values 1.18 higher than the conversion adopted in the
present paper.}  by a factor of 1.18.  The resulting
X-ray--to--optical flux ratios for the 15 CV candidates in NGC~6397
are in the range \fxfr\ $\simeq 0.5-20$, with a median value \fxfr\
$=$ 5.7.  For ease of comparison to the CV candidates in \wcen, we
have plotted the NGC~6397 CV candidates in Fig.\ 5 (small open black
triangles), assuming a distance of 2.4 kpc.  The two sets of
candidates occupy a similar part of the \lx\ vs. \fxfr\ plane, with a
similar range of flux ratios and a similar maximum X-ray luminosity.
The only apparent difference is the presence of fainter CVs in the
NGC~6397.  Whether \wcen\ also harbors CVs this faint is not yet
known, as they are below the detection limit of the \Chandra\ imaging
reported by HCD09.

\begin{figure}
%\plotone{f8.eps} 
\centerline{
\includegraphics[scale=0.5]{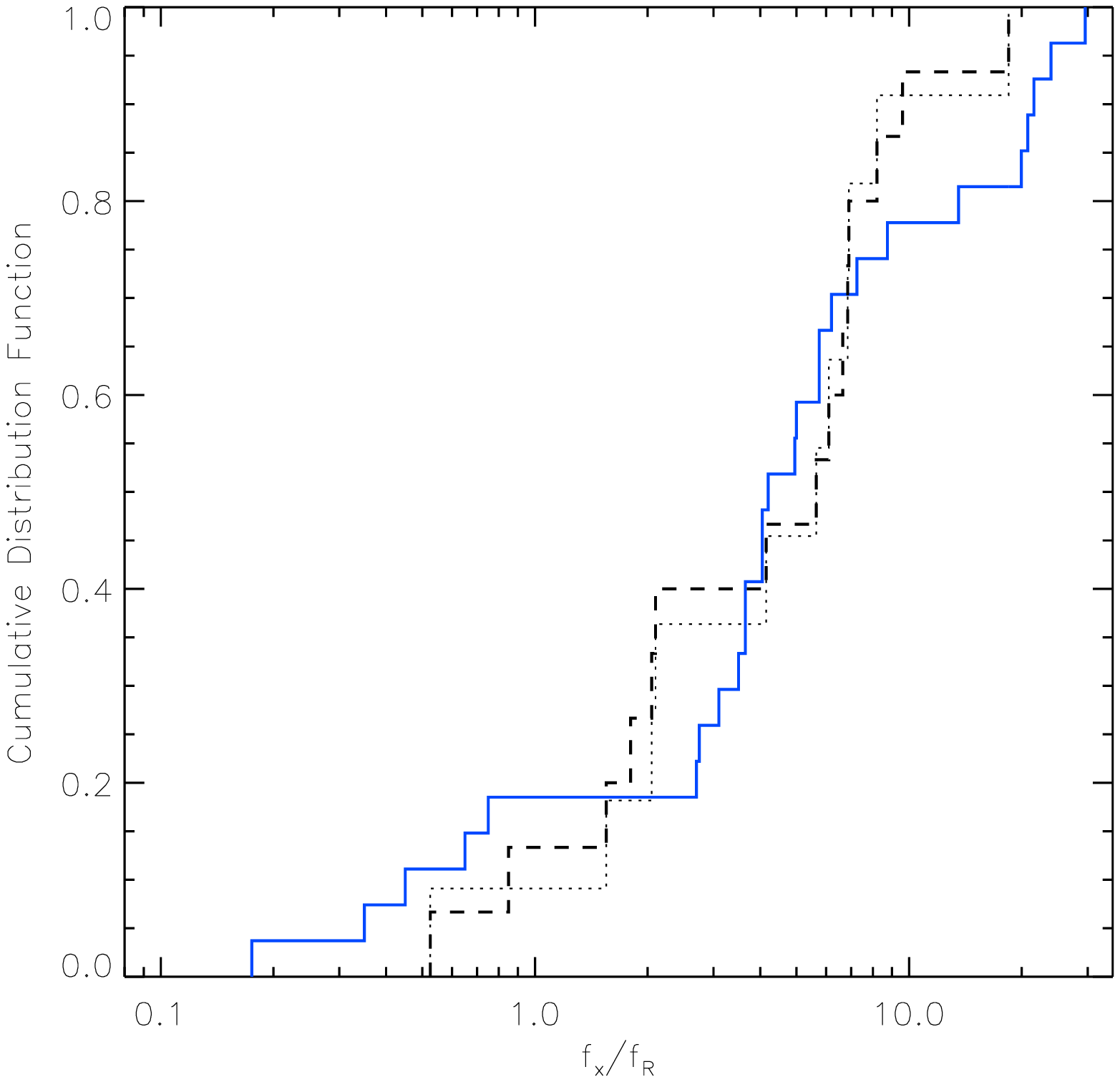}
}
\figcaption{Cumulative
  distribution of X-ray--to--optical flux ratio for CVs in
  \wcen\ vs.\ NGC~6397.  Solid blue line: CV candidates in \wcen;
  dashed black line: CV candidates in NGC~6397 that are bright enough
  to have been detected in \wcen; dotted black line: all CV candidates in
  NGC~6397.}
\end{figure}

As a quantitative test of the similarity or difference between the CV
populations in \wcen\ and NGC~6397, we use the KS test to compare
cumulative distributions of \fxfr\ ratios (see Fig.\ 8).  For
NGC~6397, we plot distributions both for the entire population of 15
candidate CVs and for the subset of 11 candidates that are bright
enough in both X-rays and optical to have been detected, at least in
principle, if they had been in \wcen.  The KS test shows that
distributions of \fxfr\ ratios are consistent with having been drawn
from the same parent population (KS probability 0.94 when comparing
\wcen\ CV candidates to the 11 brightest candidates in NGC~6397, and
0.71 when comparing to all NGC~6397 candidates).  The Anderson-Darling
test, which is more sensitive to deviations near the beginning or end
of the distributions, also reveals no significant differences: the
P-value is 0.53 when comparing \wcen\ CV candidates to those CVs in
NGC~6397 that could have been detected in \wcen, and 0.45 when
comparing to the full set of NGC~6397 CVs.  To the extent that
different \fxfr\ ratios characterize the different subclasses of CVs
(see, e.g., Verbunt et al.\ 1997, Heinke et al.\ 2008, Ag\"{u}eros et
al.\ 2009), we find no evidence that the CVs in the two clusters have
a different class make-up.  Thus, to the extent that it can be
discerned from a small number of \fxfr\ ratios, the different
formation histories of the CVs in the two clusters (see below) do not
seem to have a significant effect on the types of CVs that are
created.

We have also compared the flux ratios for CVs in \wcen\ to those
reported by Edmonds et al.\ (2003b) for the CVs in 47~Tuc.  The median
\fxfopt\ ratio they report for 17 CVs in that cluster is 1.2---a
factor of 3.5 times lower than the median value for CV candidates in
\wcen, and 4.8 times lower than the median for those in NGC~6397.
These apparent differences are not due to the different definition of
optical fluxes adopted here vs.\ by Edmonds et al.  By directly
comparing our \fxfr\ ratios to the \fxfopt\ ratios reported for
47~Tuc, we are effectively assuming that the CVs in \wcen\ and
NGC~6397 have $V-R = 1.15$.  If the CVs are bluer (as is likely given
their \br\ colors), then the differences in median flux ratios would
only increase.  It is unclear, however, what the significance of these
apparent differences is.  The faintest counterparts in the 47~Tuc CV
sample, which is based on WFPC2 imaging, have absolute magnitudes of
M$_V$ \about 10 (Edmonds et al.\ 2003a).  This is to be compared to
limits of M$_{625}$ $=$ 12.6 and 13.9 in \wcen\ and NGC~6397 CV
samples, respectively, both of which made use of the higher-resolution
ACS/WFC instrument.  Among the CV candidates in \wcen\ (though not in
NGC~6397), there is a trend toward higher \fxfr\ ratios for fainter
sources (e.g., median \fxfr\ $=$ 2.8 for sources with M$_{625}$ $<$ 10
vs.\ 6.8 for sources with M$_{625}$ $>$ 10).  If a similar trend were
present among 47~Tuc CVs, then the fact that only relatively bright
CVs are present in the sample could account for at least part of the
difference.  Interestingly, Edmonds et al.\ (2003b) show that even the
47~Tuc CVs have higher \fxfopt\ ratios than field CVs reported by
Verbunt et al.\ (1997), and a combination of X-ray luminosities and
optical fluxes that are hard to reconcile with any known class of
field CV.  These comparisons should be interpreted with caution,
however, given the difficulty of intercomparing CV samples collected
using a wide variety of discovery methods (see Knigge [2011] for a
discussion).

While the number of CV candidates in \wcen\ is among the largest yet
observed in a globular cluster, it is nevertheless quite small in
comparison to the enormous mass of the cluster, which is estimated at
4 \x\ 10$^6$ \msun\ (Pryor \&\ Meylan 1993).  Considering that the
ACS/WFC mosaic extends just beyond the half-mass radius, 27 CVs
implies \about 1.4 \x\ 10$^{-5}$ CVs per solar mass in \wcen.  For
comparison, the space density of CVs in the solar neighborhood is
\about 10$^{-5}$ pc$^{-3}$ (Patterson 1998; Pretorius et al.\ 2007),
or 10$^{-4}$ \msun$^{-1}$ considering the local mass density of \about
0.1 \msun\ pc$^{-3}$.  Even if the current census of CVs represents
only half of the true population (e.g., due to the current X-ray
detection limit and/or crowding precluding the recovery of at least
some optical counterparts), the frequency of CVs per unit mass in
\wcen\ would still be a factor of \simgt 3 lower than in the field.
While the space density of CVs is uncertain, this result suggests
that, to the extent that globular clusters both destroy CVs (and their
progenitors) and create them (Davies 1997, Ivanova et al.\ 2006,
Pooley \&\ Hut 2006), destruction may be the dominant process in
\wcen.

The opposite is likely to be true of NGC~6397.  Pooley \&\ Hut (2006)
have shown that dynamical process dominate the production of CVs in
dense clusters like NGC~6397.  A comparison of the numbers of CV
candidates in NGC~6397 vs.\ \wcen\ (counting only CVs bright enough to
have been detected in both X-rays and optical in either cluster) shows
that NGC~6397 contains nearly 7 times more CVs per unit mass than
\wcen\ (which is \about 16 times more massive than NGC~6397; Pryor \&\
Meylan 1993).  This finding further supports the view that NGC~6397 is
able to generate new CVs by dynamical processes.

The contrast in formation histories of the CVs in the two clusters may
help to explain one notable difference between the two samples.  While
the numbers of the most luminous CVs are similar in both clusters
(e.g., 5 each in the decade between \lx\ $= 2$\x $10^{31} - 2$\x
$10^{32}$ \ergs), \wcen\ contains a factor of \about 3 more faint CV
candidates in the decade from \lx\ $= 10^{30}-10^{31}$ \ergs\ (see
Fig.\ 5).  This is despite what is likely to be a lower rate of
recovery of faint optical counterparts in \wcen\ due to crowding (1.7
million stars in 9 WFC pointings in \wcen\ [Anderson \&\ van der Marel
2010] vs.\ 25,000 in NGC~6397 in one WFC pointing [Strickler et al.\
2009]).  Since CVs evolve toward fainter luminosities as they age
(e.g., Patterson 1984; Townsley \&\ G\"{a}nsicke 2009; Knigge,
Baraffe, \&\ Patterson 2011), the larger proportion of faint CVs in
\wcen\ could be indicative of a population that is older on average.
In qualitative terms at least, this is just as might be expected if
\wcen's population is dominated by CVs of primoridal origin while CVs
in NGC~6397 have been produced in dynamical interactions over much of
its lifetime.

% SECTION 4.2

\subsection{Active binaries on or near the anomalous RGB/SGB} 

An unexpected result of our search for optical counterparts of X-ray
sources in and toward \wcen\ is the discovery of a population that
appears to be associated with the anomalous giant and subgiant
branches in the cluster (see \S 3.4).  The X-ray luminosities of these
stars are in the range \about $2-40$ \x\ $10^{30}$ \ergs, which is
significantly higher than would be expected for a population of old
single stars.  It is however characteristic of RS~CVn-type binary
stars, in which a giant or subgiant star is rotating fast as a result
of tidal locking with a binary companion in a few-day orbit (Dempsey
et al.\ 1993).  The soft X-ray colors of these stars (see Fig.\ 5) are
also typical of such coronal sources (cf.\ Heinke et al.\ 2005).
Spectra are needed to determine whether these stars have the
radial-velocity variability characteristic of such binaries, and
whether their metallicities make them members of the anomalous RGB/SGB
population in the cluster.  In the meantime, given the X-ray evidence,
we assume that they are indeed binaries, and consider the implications
if they are (or are not) RGB/SGB-a members.

If these stars are members of the anomalous giant and subgiant
populations in \wcen, it would suggest that this relatively metal-rich
population is somehow able to produce binaries much more efficiently
(by a factor of 5 to 15---see \S 3.4) than the intermediate and
low-metallicity populations that dominate the cluster.  Alternatively,
the binary fraction could be similar among the different populations
if the binaries in the anomalous RGB/SGB-a population were on average
more luminous in X-rays (e.g., as a result of increased coronal
activity), such that a larger fraction were above the \Chandra\
detection limit.  The anomalous RGB/SGB-a population in \wcen\ (whose
corresponding main sequence has now also been identified---see Bellini
et al.\ 2010) is a factor of \about 10 more rich in metals than the
dominant population in the cluster ([Fe/H]$\simeq -0.6$ to $-0.8$ vs.\
[Fe/H]$=-1.7$; Sollima et al.\ 2005, Johnson \a\ Pilachowski 2010).
It is thought to be the last in a series of discrete star formation
episodes early in the life of \wcen, producing increasingly high metal
abundances as a result of self-enrichment.  A factor of \about 5--15
enhancement in the binary fraction in this population would be
surprising considering that the frequency of field binaries shows no
dependence on metallicity (Latham et al.\ 2002, Carney et al.\ 2005).
On the other hand, metallicity is known to have a significant effect
on the frequency of another class of binaries in globular clusters:
LMXBs are three times more likely to be found in metal-rich clusters
than in metal-poor clusters (Grindlay 1993, Bellazzini et al.\ 1995).
While the cause for this difference is still debated (see Ivanova et
al.\ 2006 and references therein), a second case of enhanced binary
fraction in a metal-rich cluster population would be quite
interesting.  We note that recent theoretical work by Vesperini et
al.\ (2011) suggests that significant differences could exist between
the binary star populations in first-generation vs.\ second-generation
stars in globular clusters with multiple stellar populations.
However, their study finds an opposite effect to what is observed here
(assuming these stars are members of the anomalous population), with
binaries being more abundant among the first generation of stars.

If the metallicities of these stars reveal instead that they are
members either of the metal-poor population or one of the
intermediate-metallicity populations in \wcen\ (Sollima et al.\ 2005),
then their location in the CMD carries a different significance.  In
this case they would add to a growing number of globular and open
clusters in which small number of red stragglers have been found
(e.g., Belloni et al.\ 1998, Albrow et al.\ 2001, Kaluzny 2003,
Edmonds et al.\ 2003a, Bassa et al.\ 2004, Bassa et al.\ 2008, Geller
et al.\ 2008, Huang et al.\ 2010, Cohn et al.\ 2010, Platais et al.\
2011).  Red stragglers are loosely defined to be stars that lie
redward of the turnoff.  Typically they appear below the subgiant
branch, but there are exceptions to this (see below).  The term was
coined by Albrow et al.\ (2001) to describe a group of six variable
stars that appeared on the red side of the main-sequence turnoff in
their HST/WFPC2 study of 47~Tuc.  Five of these are likely BY~Dra-type
binary stars with periods in the range 4.9-9.2 days; the sixth is a
CV.  Three additional red stragglers were identified by Edmonds et
al.\ (2003a) as optical counterparts of \Chandra\ sources.  All but
one of the red stragglers in 47~Tuc have \Chandra\ counterparts
(Edmonds et al.\ 2003a, Heinke et al.\ 2005).  Like the stars in
question in \wcen, the X-ray luminosities and soft X-ray colors of
these stars (with the exception of the known CV) are indicative of
coronal sources, consistent with their being binaries containing
chromospherically active stars.

Red stragglers are particularly intriguing because they cannot be
explained by the simple superposition of two normal stars.  In
general, some sort of mass transfer, which can take a star out of
thermal equilibrium at least temporarily, has been invoked to make
sense of them.  The most detailed study to date of stars redward of
the turnoff is that of two stars that appear below the subgiant branch
in the old open cluster M67 (Mathieu et al.\ 2003).  Proper motions
and radial velocities both point to their being probable cluster
members.  Dubbed ``sub-subgiants'' by Belloni et al.\ (1998),
spectroscopic analyses reveal that both are binaries, with periods of
2.8 and 18.4 days, respectively, and have the strong Ca II H and K
emission line cores characteristic of coronal sources as well as \ha\
in emission (Pasquini \&\ Belloni 1998, van den Berg et al.\ 1999).
Their X-ray luminosities are also typical of such systems (Belloni et
al.\ 1998).  Mathieu et al.\ (2003) conduct an exhaustive analysis of
all available data and conclude that it is likely that both binaries
have experienced mass transfer and/or significant dynamical
interactions with other stars (see also Hurley et al.\ 2001), but are
unable to fully explain their locations \about 1 magnitude below the
subgiant branch.  Clearly there is more to be learned from these
enigmatic stars.

One important step will be to obtain more complete and unbiased
samples than currently exist.  Until recently nearly all red
stragglers known in globular clusters were identified either as
counterparts to X-ray sources or as photometric variables.  Both of
these detection methods are strongly biased toward finding binaries.
They are also likely to select only the more extreme members of a
larger class.  Even the region in the CMD that such stars can occupy
is unclear at present.  For example, while the terms ``red straggler''
and ``sub-subgiant'' have generally been considered interchangeable,
it appears that not all such stars that arguably belong in this class
are fainter than subgiants.  In the open cluster NGC~188, two cluster
members (one of which has been shown to be binary) are adjacent to the
base of the red giant branch (Geller et al.\ 2008).  In \wcen, two or
three of the red straggler candidates we identify here are also
brighter than subgiants (depending on which population of subgiants is
chosen as a reference).

Proper motion measurements can ameliorate this situation by providing
unbiased samples in globular clusters, as has been done for open
clusters.  The recent study of red stragglers in \wcen\ that makes use
of ground-based proper motions measurements by Bellini et al.\ (2009)
is instructive.  Rozyczka et al.\ (2012) identify 13 red stragglers in
\wcen\ (the largest number yet uncovered in any study), all of which
have colors significantly redder even than the RGB/SGB-a sequences in
the cluster (and one of which is brighter than the subgiant branch).
In the absence of both proper motion and radial velocity measurements
suggesting otherwise, these stars would likely have been dismissed as
unrelated to the cluster.  Interestingly, Rozyczka et al.\ (2012) find
evidence that well over half are binaries, and surmise that they all
may be.  Given the extreme that \wcen\ represents among globular
clusters in many regards (including the recent and suprising discovery
of a population of faint and very red main-sequence stars; King et
al.\ 2012), it remains to be seen whether such extreme red stragglers
are commonplace in globulars.

In comparing red stragglers in globular clusters vs.\ open clusters,
it is notable that the numbers seen in globulars are not in general
much greater than the numbers seen in open clusters, despite the
vastly different total numbers of stars they contain.  In \wcen, with
an estimated mass of 3 \x\ 10$^6$ \msun, 20 are now known.  Scaling
these numbers to M67, whose mass is more than 3 orders of magnitude
lower (Fan et al.\ 1996), one would not expect to find any such stars.
Yet two are known.  NGC~6791, an open cluster with eight known red
stragglers (Kaluzny 2003, Platais et al.\ 2011), is an even more
extreme example.  Despite the incompleteness of the current samples,
it seems very likely that open clusters have a sigificantly higher
specific frequency of red stragglers than globular clusters.

% SECTION 4.3

\subsection{Where are the main-sequence binaries?}

By comparison to the large numbers of CV candidates we have identified
among the \Chandra\ sources, the numbers of potential main-sequence
binaries in the form of BY~Dra stars is very small.  This is in stark
contrast to the situation in NGC~6397, in which a search using the
same camera and filters on \HST\ revealed nearly three times as many
active binaries as CVs (42 vs.\ 15; Cohn et al.\ 2010).  A smaller
yield is to be expected in \wcen\ due to a higher degree of crowding
and its larger distance (making the limit of the present search \about
1.5 magnitudes brighter than in NGC~6397).  Taking these factors into
account, we estimate that \about 10 of the NGC~6397 ABs, if present in
\wcen, should have been detectable in our optical data.  Finding only
3 BY~Dra stars is even more surprising in view of the fact that
\wcen\ is a factor of \about 40 times more massive than NGC~6397
(Harris 1996), and that the present mosaic encompasses \about 1.7
million measured stars (vs.\ \about 25,000 in the NGC~6397 dataset).
If the two clusters harbor similar fractions of close MS binaries one
might have expected hundreds of ABs in \wcen.

The critical difference between the two searches is not the optical
data, but the X-ray data.  Whereas the limiting X-ray luminosity in
the existing \wcen\ imaging is \lx \about 10$^{30}$ \ergs, the limit
is \simlt 10$^{29}$ \ergs\ in NGC~6397---more than an order of
magnitude lower.  Nevertheless, if \wcen\ does harbor as large a
population of BY~Dra stars as NGC~6397 (per unit mass), it is still
surprising to have found so few, considering that Dempsey et al.\
(1997) find that about 15\% of field BY~Dra binaries have \lx $>$
10$^{30}$ \ergs\ (see their Fig.\ 4).  On the other hand, of the $>$30
ABs in NGC~6397 that lie in the part of the CMD associated with
main-sequence binaries, not a single one is bright enough in X-rays to
have been detected in the \wcen\ \Chandra\ imaging (see Fig.\ 6 of
Cohn et al.\ 2010).  This is interesting in itself, as it suggests
that there may be significant differences between the populations of
short-period MS binaries in globular clusters vs.\ the field.  Deeper
\Chandra\ imaging is clearly needed to determine whether there are
significant differences between the short-period MS binary populations
in \wcen\ vs.\ NGC~6397.

% SECTION 5

\section{Summary and Conclusions}

We have used ACS/WFC to identify optical counterparts for 59 \Chandra\
sources in and toward Omega Centauri.  Among the sources likely to be
associated with the cluster are 27 candidate cataclysmic variables, a
blue straggler, a qLMXB, and three possible BY~Draconis-type binaries.
In addition, we find 7 giants and subgiants whose locations in a
color--magnitude diagram suggest either that they are red stragglers
or that they are members of the most metal-rich RGB/SGB population in
the cluster.

The frequency of CVs in \wcen\ appears to be lower than the frequency
of CVs in the Galactic field, by a factor of \about 3 if we assume
that half of \wcen\ CVs have yet to be discovered.  This suggests that
the majority of binaries that would give rise to CVs in the field are
destroyed in the cluster environment.  Alternatively, the primordial
binary fraction in \wcen\ may have been lower than that in the field.
We have also compared the properties of the CVs in \wcen\ to the CVs
in NGC~6397.  Both are X-ray-selected samples, with optical IDs
obtained using the same method.  CVs are 7 times more frequent per
unit mass in NGC~6397 than in \wcen, and are likely to have been
formed primarily through dynamical interactions.  We find no
measurable difference in the distribution of X-ray--to--optical flux
ratios for CVs in the two clusters.  To the extent that
X-ray--to--optical flux ratios are indicators of CV subtype, we find
no indication that the types of CVs present in the two clusters
differ, despite their constrasting formation histories.

Globular clusters have the potential to test CV evolutionary theory by
providing samples of CVs all at the same relatively well-known
distance---albeit with the added complication of cluster dynamics.
Among the CVs in \wcen\ is a faint group with absolute magnitudes
similar to those of field CVs identified in the Sloan Digital Sky
Survey, whose orbital periods are near the theoretical minimum for CVs
(G\"{a}nsicke et al.\ 2009).  A similarly faint population of CVs was
identified in NGC~6397 by Cohn et al.\ (2010).  Thus, both clusters
for which sufficiently sensitive observations have been made are found
to contain significant numbers of these faint systems.  This provides
qualitative evidence in support of the theory, which predicts a pile
up of old, faint CVs near the period mininimum.  However, orbital
periods are needed before more definitive conclusions can be drawn.
Meaningful CV luminosity functions will also require determining the
extent to which existing samples are incomplete due to the effects of
crowding in optical imaging.  In the case of \wcen, deeper X-ray
observations are also needed to sample the faintest systems.

One notable difference between the CV populations now known in \wcen\
vs.\ NGC~6397 is the relative numbers of bright vs.\ faint CVs.  The
proportion of faint systems is a factor of \about 2-3 times higher in
\wcen\ than in NGC~6397, which suggests that the CVs in \wcen\ may be
older on average than those in NGC~6397.  This supports the view that
\wcen's population is dominated by CVs that derive from primordial
binaries, while NGC~6397 is continually manufacturing new compact
binaries through dynamical interactions.

While nearly 50\% of the optical counterparts we have identified
appear to be compact binaries, only about 5\% have characteristics
indicative of BY~Dra stars.  This is in contrast to NGC~6397, in which
active main-sequence binaries outnumber CVs by almost three to one
(Cohn et al.\ 2010).  These faint X-ray sources provide a valuable
window into populations of short-period main-sequence binaries in
globular clusters.  Deeper \Chandra\ imaging of \wcen\ is needed to
determine whether there are real differences between the relative
numbers of these two classes of binaries in the cluster.  The paucity
of BY~Dra candidates in the present study may simply be a consequence
of the \about 10$^{30}$ \ergs\ limiting luminosity of the existing
\Chandra\ study (HCD09).

Seven of the optical counterparts have magnitudes and colors that
place them on or near the anomalous giant and subgiant branches in
\wcen.  The X-ray properties of these stars suggest that they may be
RS~CVn-type binaries.  If the apparent association between these stars
and the RGB/SGB-a stars is real, then the frequency of binaries in
this metal-rich population is enhanced by a factor of five relative to
the other giant and subgiant populations in the cluster.
Spectroscopic observations are needed to determine whether or not
these stars have metallicities that indicate membership in the
RGB/SGB-a population.

If these stars are not members of \wcen's most metal-rich population,
then they lie in a region of the CMD that cannot be explained by
single-star evolution.  In this case, they add to a growing number of
red stragglers that have been identified in \wcen, making it the
cluster with the largest such population yet known.

\vspace{-3mm}
\acknowledgments

We gratefully acknowledge discussions with Craig Heinke, Aaron Geller,
Haldan Cohn and Phyllis Lugger, and thank the anonymous referee for
helpful comments that improved the manuscript.  This work is based on
observations with the NASA/ESA Hubble Space Telescope and was
supported by NASA grant HST-GO-9442 from the Space Telescope Science
Institute, which is operated by the Association of Universities for
Research in Astronomy, Incorporated, under NASA contract NAS5-26555.
Support was also provided by Chandra Award Number GO0-1040A issued by
the Chandra X-ray Observatory Center, which is operated by the
Smithsonian Astrophysical Observatory for and on behalf of the
National Aeronautics Space Administration under contract NAS8-03060.

\clearpage

\begin{deluxetable}{lccccccccc}
\vspace{-1.5cm}
\tablefontsize{\scriptsize}
\tablewidth{0pt}
\tablecaption{Candidate Optical Counterparts: Astrometry} 
\tablehead{\colhead{X-ray$^a$} & \colhead{Optical$^b$} & \colhead{x} &  \colhead{y} & \colhead{offset$^c$} & \colhead{quality$^d$} & \colhead{type$^e$} & \colhead{RA} & \colhead{Dec} & \colhead{cluster offset$^f$} \\
\colhead{ID} & \colhead{ID\#} & \colhead{(pix)} & \colhead{(pix)} & \colhead{(pix)} & \colhead{} & \colhead{} & \colhead{(J2000)} & \colhead{(J2000)} & \colhead{(arcsec)} }

\startdata
12a  &     1486  &  207.5  &  201.9  &    4.9   &  1  & CV &                         13:26:48.651  &  $-$47:27:44.82  &   63  \\
13a  &     9205  &  205.3  &  201.8  &    4.9   &  0  & CV &                         13:26:53.513  &  $-$47:29:00.38  &   65  \\
13b  &     1398  &  207.8  &  194.9  &   11.8  &  1  & RGB/SGB-a &                 13:26:50.532  &  $-$47:29:18.15  &   46  \\
13c  &     1408  &  206.4  &  200.8  &    5.8   &  1  & CV &                         13:26:52.135  &  $-$47:29:35.63  &   69  \\
13f  &     3001  &  205.0  &  201.8  &    4.9   &  1  & CV? &                        13:26:45.980  &  $-$47:29:16.63  &   32  \\
14a  &     8018  &  200.2  &  208.1  &    6.1   &  1  & BYDra &                      13:26:45.770  &  $-$47:28:59.19  &   19  \\
21b  &     8001  &  206.7  &  204.9  &    1.8   &  1  & fbCV &                       13:26:35.341  &  $-$47:27:59.15  &  129  \\
21c  &     9121  &  212.3  &  209.2  &    6.7   &  1  & \ha-only &                   13:26:36.888  &  $-$47:27:45.72  &  121  \\
""   &     1310  &  212.4  &  203.9  &    6.9    &  1   &  \ha-only             &  13:26:36.862  &  $-$47:27:45.75  &  121  \\
21d  &     4017  &  206.7  &  213.7  &    7.1  &  1  & \ha-only &                  13:26:38.308  &  $-$47:27:40.36  &  112  \\
""   &     1350  &  202.9  &  210.3  &    4.9    &  1   &  blue-only                       &  13:26:38.292  &  $-$47:27:40.58  &  112  \\
22a  &     1074  &  206.3  &  204.6  &    2.0   &  0  &FGND &                       13:26:48.299  &  $-$47:26:41.21  &  125  \\
22c  &     1347  &  206.3  &  207.9  &    1.3   &  0  &fbCV &                       13:26:52.687  &  $-$47:27:13.41  &  108  \\
22d  &     4003  &  205.7  &  201.3  &    5.3   &  1  &BS &                         13:26:58.746  &  $-$47:27:28.93  &  140  \\
22e  &     1340  &  204.2  &  202.1  &    4.9   &  1  &RGB/SGB-a &                  13:26:59.926  &  $-$47:28:09.68  &  133  \\
22f  &     1287  &  205.2  &  195.3  &   11.3   &  1  &AGN &                        13:26:58.809  &  $-$47:28:21.19  &  119  \\
""   &     1334  &  196.7  &  202.4  &   10.3    &  1   &   blue-only                      &  13:26:58.847  &  $-$47:28:21.56  &  120  \\
23b  &     1187  &  207.2  &  195.7  &   11.0  &  1  & CV? &                       13:26:51.683  &  $-$47:30:47.33  &  128  \\
""   &     1238  &  211.1  &  202.4  &    6.5     &  1  &  blue-only                        &  13:26:51.714  &  $-$47:30:47.09  &  128  \\
23c  &  5000036  &  211.5  &  197.7  & 10.4  &  1  &   AGN &                       	13:26:48.051  &  $-$47:30:14.36  &   88  \\
24a  &     1413  &  205.2  &  204.4  &    2.4    &  1 & blue-only                &  13:26:44.466  &  $-$47:30:06.03  &   84  \\
24c  &     1217  &  207.3  &  204.0  &    2.9   &  0  & CV &                         13:26:38.417  &  $-$47:30:36.75  &  141  \\
24e  &     1034  &  206.1  &  217.4  &   10.8  &  1  & BYDra &                     13:26:36.852  &  $-$47:30:11.55  &  135  \\
24f  &  5000592  &  201.7  &  209.7  &  5.4  &  1  &   RGB/SGB-a &                 	13:26:37.287  &  $-$47:29:42.93  &  115  \\
24g  &     6001  &  207.5  &  205.5  &    1.8  &  1  & AGN? &                      13:26:34.387  &  $-$47:29:55.71  &  147  \\
31a  &     7005  &  209.5  &  199.1  &    8.2  &  0  & fbCV &                      13:26:29.356  &  $-$47:28:13.21  &  184  \\
31b  &     9002  &  204.6  &  208.6  &    2.5  &  1  & AGN &                       13:26:31.391  &  $-$47:28:01.53  &  166  \\
32a  &     9001  &  209.1  &  207.8  &    3.2  &  0  & fbCV &                      13:26:46.353  &  $-$47:25:18.45  &  208  \\
32c  &     1082  &  207.1  &  210.1  &    3.6  &  1  & \ha-only &                  13:26:55.907  &  $-$47:26:02.18  &  186  \\
32f  &     7003  &  203.8  &  201.5  &    5.6  &  1  & RGB/SGB-a &                 13:27:05.331  &  $-$47:28:08.78  &  187  \\
33c  &     1248  &  210.0  &  205.5  &    4.1  &  1  & blue-only &                 13:27:03.620  &  $-$47:28:57.87  &  166  \\
33d  &     1268  &  197.2  &  206.0  &    8.9  &  0  & \ha-only &                  13:27:01.461  &  $-$47:29:25.17  &  149  \\
33e  &     1281  &  199.9  &  211.3  &    7.8  &  0  & CV? &                       13:27:00.974  &  $-$47:30:04.72  &  159  \\
""   &     3128  &  198.8  &  215.4  &   11.4  &  1  &  CV? &                          13:27:00.995  &  $-$47:30:04.75  &  159  \\
33h  &     1110  &  207.3  &  195.9  &   10.8 &  1  &  blue-only &                13:26:55.066  &  $-$47:31:13.69  &  167  \\
33j  &     3001  &  199.0  &  213.0  &    9.6   &  0  &fbCV &                       13:26:49.621  &  $-$47:31:24.83  &  160  \\
33l  &     1117  &  207.1  &  200.8  &    5.9   &  1  &AGN? &                       13:26:48.731  &  $-$47:31:25.28  &  159  \\
33m  &     1112  &  214.3  &  207.0  &    8.2 &  1  &  CV? &                      13:26:46.491  &  $-$47:31:40.75  &  174  \\
34b  &     1204  &  200.2  &  208.3  &    6.1  &  0  & RGB/SGB-a &                 13:26:37.440  &  $-$47:30:53.34  &  161  \\
41a  &      637  &  205.6  &  205.3  &    1.4   &  1  &\ha-only &                   13:26:24.423  &  $-$47:26:57.72  &  255  \\
41d  &      930  &  207.8  &  201.6  &    5.3   &  1  &CV &                         13:26:28.651  &  $-$47:26:27.29  &  234  \\
41g  &     6005  &  203.3  &  212.6  &    6.6  &  1  & AGN &                       13:26:37.460  &  $-$47:24:29.93  &  275  \\
41h  &      915  &  208.3  &  218.3  &   11.9  &  0  & RGB/SGB-a &                 13:26:43.958  &  $-$47:24:42.63  &  246  \\
42c  &     1015  &  211.8  &  208.8  &    6.1  &  0  & RGB/SGB-a &                 13:27:09.652  &  $-$47:27:28.85  &  240  \\
43c  &     1347  &  205.6  &  200.0  &    6.6  &  1  & RGB/SGB-a &                 13:27:06.910  &  $-$47:30:09.42  &  215  \\
43e  &     1029  &  203.4  &  196.9  &   10.1 &  0  &  BYDra &                    13:27:03.426  &  $-$47:30:56.33  &  209  \\
43f  &     2001  &  205.5  &  203.6  &    3.1  &  0  & FGND &                      13:26:56.047  &  $-$47:32:02.16  &  215  \\
43h  &      914  &  202.5  &  202.7  &    5.3   &  1  &CV? &                        13:26:49.584  &  $-$47:32:12.83  &  207  \\
44a  &      833  &  207.4  &  199.1  &    7.6   &  1  &blue-only &                  13:26:44.102  &  $-$47:32:31.46  &  227  \\
44c  &     7003  &  210.2  &  209.5  &    5.0  &  0  & fbCV &                      13:26:23.641  &  $-$47:30:43.83  &  266  \\
""   &      798  &  200.8  &  216.1  &   10.9   &  1  & AGN? &                           13:26:23.678  &  $-$47:30:44.27  &  266  \\
44d  &      896  &  209.9  &  204.6  &    4.3   &  0  &CV? &                        13:26:22.858  &  $-$47:30:09.14  &  260  \\
44e  &      788  &  206.3  &  207.4  &    0.8   &  0  &qLMXB &                      13:26:19.796  &  $-$47:29:10.51  &  279  \\
51a  &  5000018  &  214.7  &  207.2  & 8.6  &  1  &    blue-only &                 	13:26:31.315  &  $-$47:24:39.27  &  295  \\
51d  &      670  &  208.3  &  202.2  &    4.9  &  0  & \ha-only &                  13:26:40.994  &  $-$47:24:02.22  &  291  \\
51e  &      823  &  207.1  &  215.3  &    8.8  &  1  & FGND &                      13:26:44.764  &  $-$47:23:33.59  &  313  \\
""   &  5000005  &  215.4  &  204.4  &  9.6  &  1  &  FGND  &                          	13:26:44.708  &  $-$47:23:33.24  &  314  \\
52b  &      738  &  210.0  &  214.3  &    8.6  &  0  & AGN &                       13:26:54.948  &  $-$47:24:08.88  &  288  \\
""   &      748  &  207.3  &  217.9  &   11.4  &  0  & AGN &                          13:26:54.967  &  $-$47:24:08.99  &  288  \\
52c  &      521  &  201.4  &  209.1  &    5.3  &  0  & AGN? &                      13:27:06.396  &  $-$47:25:38.20  &  270  \\
52d  &      796  &  204.1  &  204.7  &    2.8  &  0  & fbCV &                      13:27:14.923  &  $-$47:27:43.54  &  287  \\
54b  &     5008  &  206.3  &  207.4  &    0.8  &  1  & fbCV &                      13:26:42.452  &  $-$47:33:09.20  &  267  \\
54d  &      617  &  206.2  &  207.7  &    1.1  &  0  & AGN &                       13:26:25.127  &  $-$47:32:27.40  &  314  \\
""   &      637  &  206.1  &  214.2  &    7.6  &  0  & FGND &                          13:26:25.159  &  $-$47:32:27.35  &  314  \\
54e  &     5006  &  202.5  &  203.9  &    4.5  &  1  & fbCV &                      13:26:25.384  &  $-$47:31:40.97  &  281  \\
54g  &      747  &  207.1  &  209.8  &    3.4  &  1  & AGN &                       13:26:20.034  &  $-$47:30:15.18  &  289  \\
""   &      715  &  212.9  &  200.3  &    9.3  &  1  & \ha-only &                          13:26:19.986  &  $-$47:30:14.97  &  290  \\
54h  &      734  &  206.7  &  206.4  &    0.6  &  1  & CV &                        13:26:20.366  &  $-$47:30:02.94  &  282  \\
62b  &      538  &  202.2  &  212.8  &    7.3  &  0  & AGN &                       13:27:08.011  &  $-$47:23:33.86  &  377  \\
\enddata
\tablecomments{Astrometric properties of the candidate optical counterparts to the {\it Chandra} X-ray sources, sorted by X-ray ID (roughly the distance from the cluster center).
$^a$X-ray ID from HCD09.
$^b$Optical ID\# from DAOPHOT (see \S 2.3). 
$^c$Offset between the X-ray source and optical counterpart positions. The X-ray source position in physical coordinates is (x,y) = (206.1,206.6) in each patch after boresite correction (\S 2.1). The optical counterpart position in physical coordinates appears in the previous two columns. 
$^d$Estimated quality of the photometry; 0=excellent, 1=good (see \S 2.3 for details).
$^e$Abbreviation for the most likely source classification: quiescent low mass X-ray binary (qLMXB), cataclysmic variable (CV), possible CV (CV?), faint blue CV (fbCV), BY Draconis-type binary (BYDra), blue straggler (BS), star located on or near anomalous subgiant or giant branch in CMD (RGB/SGB-a), \ha-bright source with no measurable blue excess (\ha-only), foreground star (FGND), background active galactic nucleus (AGN), possible AGN (AGN?), blue source with no measurable \ha-excess (blue-only). Refer to Table 2 and \S 3 for photometric properties and a discussion of these classifications. 
$^f$Offset between optical source position and cluster center, (RA,Dec) = (13:26:47.24,$-$47:28:46.45) from Anderson \& van der Marel (2010).
}
%\label{astrom_tab}
\end{deluxetable}

\begin{deluxetable}{lccccccccc}
\vspace{-1.0cm}
\tablefontsize{\scriptsize}
\tablewidth{0pt}
\tablecaption{Candidate Optical Counterparts: Photometry} 
\tablehead{\colhead{X-ray$^a$} & \colhead{\r} & \colhead{\br} & \colhead{\hr} & \colhead{color$^b$} & \colhead{\ha$^c$} & \colhead{quality} & \colhead{type} & \colhead{\fratio$^d$} & \colhead{${L}_X$$^e$} \\
\colhead{ID} & \colhead{(mag)} & \colhead{(mag)} & \colhead{(mag)} & \colhead{} & \colhead{} & \colhead{} & \colhead{} & \colhead{} & \colhead{($10^{30}$ erg s$^{-1}$)} }

\startdata
44e  &  25.3  &  \phs 1.5  &  $-$1.2  &      blue  &    bright  & 0 & qLMXB  &  440  & 120\\

13c  &  20.9  &  \phs 0.7  &  $-$0.5  &      blue  &    bright  & 1 &   CV  &    8.8 & 140  \\
13a  &  20.0  &  \phs 1.1  &  $-$0.4  &     blue?  &    bright  & 0 &   CV  &    4.1  & 150   \\
12a  &  21.3  &  \phs 1.0  &  $-$0.3  &      blue  &    bright  & 1 &    CV  &   4.2  &  48   \\
54h  &  20.7  &  \phs 0.2  &  $-$0.1  &      blue  &    bright  & 1 &    CV  &   3.7  &  72   \\
41d  &  22.2  &  \phs 1.3  &  $-$0.5  &      blue  &    bright  & 1 &    CV  &   5.0  &  24   \\
24c  &  23.5  &  \phs 0.5  &  $-$1.0  &      blue  &    bright  & 0 &    CV  &   2.7  &   4.1   \\

33e$^f$  &  22.0  &  \phs 1.4  &  $-$0.2  &      blue  &   bright?  & 0   &      CV?  &    0.46  &  2.6  \\
43h  &  22.6  &  \phs 0.5  &  $-$0.1  &      blue  &   bright?  & 1   &      CV?  &    3.7  & 12  \\
33m  &  22.7  &  \phs 1.7  &  $-$0.4  &     blue?  &    bright  & 1  &       CV?  &   0.80  &  2.5  \\
13f  &  24.2  &  \phs 1.0  &  $-$0.7  &      blue  &   bright?  &    1  &   CV?  &   3.5  &  2.7  \\
23b$^f$  &  24.5  &  \phs 0.7  &  $-$0.6  &      blue  &   bright?  &   1  &    CV?  &  4.1  &  2.4  \\
44d  &  25.2  &  \phs 0.5  &  $-$1.0  &      blue  &   bright?  &   0   &   CV?  &    21  &  6.3  \\

22c  &  24.5  &  \phs 1.0  &  \phs\nodata  &      blue  &    \nodata  &  0  &    fbCV  &   22  & 13     \\
31a  &  24.6  &  \phs 0.4  &  \phs\nodata  &      blue  &    \nodata  &  0  &    fbCV  &   14  &  7.2     \\
21b  &  24.8  &  \phs 0.0  &  \phs\nodata  &      blue  &    \nodata  & 1  &      fbCV  &   3.1  &  1.4     \\
52d  &  24.8  &  \phs 0.2  &  \phs\nodata  &      blue  &    \nodata  & 0 &      fbCV  &    7.3  &  3.2     \\
33j  &  24.9  &  $-$0.3    &  \phs\nodata  &      blue  &    \nodata  &  0 &     fbCV  &    5.0  &  2.1     \\
54b  &  25.2  &  \phs 1.4  &  \phs\nodata  &      blue  &    \nodata  &  1 &     fbCV  &   20  &  6.2     \\
54e  &  25.9  &  \phs 1.0  &  \phs\nodata  &     blue?  &    \nodata  & 1  &      fbCV  &   5.8  &  1.0     \\
32a  &  26.3  &  $-$0.4    &  \phs\nodata  &      blue  &    \nodata  &  0  &     fbCV  &  24  &  2.6     \\
44c$^f$  &  26.3  &  \phs 0.2  &  \phs\nodata  &      blue  &    \nodata  &  0  &    fbCV  & 30  &  3.3     \\

24e  &  19.7  &  \phs 1.4  &  \phs 0.1  &      red?  &   bright?  & 1  &    BYDra  &  0.013  &  0.6 \\
43e  &  20.3  &  \phs 1.5  &  \phs 0.1  &      red?  &   bright?  & 0  &    BYDra  &  0.055  &  1.5 \\
14a  &  20.5  &  \phs 1.5  &  \phs 0.1  &      red?  &   bright?  & 1   &   BYDra  &  0.050  &  1.2 \\

22d  &  17.0  &  \phs 0.9  &  $-$0.1  &     blue?  &    bright  &   1   &    BS  &    0.007  &  3.8 \\

24f  &  15.5  &  \phs 1.5  &  \phs 0.1  &       red  &   neither  & 1 & RGB/SGB-a  &    0.0007 &   1.6    \\
22e  &  16.5  &  \phs 1.5  &  \phs 0.1  &       red  &   bright?  & 1 & RGB/SGB-a  &    0.009 &   8.0    \\
32f  &  16.8  &  \phs 1.5  &  \phs 0.1  &       red  &   bright?  & 1 & RGB/SGB-a  &    0.006 &   4.0    \\
13b  &  17.2  &  \phs 1.1  &  \phs 0.3  &       red  &   neither  & 1 & RGB/SGB-a  &    0.010 &   4.9    \\
43c  &  17.2  &  \phs 1.3  &  \phs 0.0  &       red  &   bright?  & 1 & RGB/SGB-a  &    0.004 &   1.8    \\
34b  &  17.5  &  \phs 1.3  &  $-$0.1  &       red  &   neither  & 0 & RGB/SGB-a  &      0.075 &  28    \\
42c  &  17.9  &  \phs 1.2  &  \phs 0.2  &       red  &   neither  & 0 & RGB/SGB-a  &    0.005 &   1.4    \\
41h  &  18.1  &  \phs 1.1  &  \phs 0.2  &       red  &   neither  & 0 & RGB/SGB-a  &    0.006 &   1.4    \\

21d$^f$  &  21.1  &  \phs 1.7  &  $-$0.1  &   neither  &   bright?  & 1 &  \ha-only         &   0.39  &  5.3 \\
32c  &  21.1  &  \phs 1.6  &  \phs 0.0  &   neither  &   bright?  & 1  &  \ha-only      &   0.18  &  2.4 \\
33d  &  21.8  &  \phs 2.0  &  \phs 0.0  &   neither  &   bright?  & 0  &  \ha-only      &   0.68  &  4.8 \\
51d  &  22.9  &  \phs 2.3  &  $-$0.1  &   neither  &   bright?  &  0 &  \ha-only        &    2.8  &  7.1 \\
41a  &  24.5  &  \phs 2.5  &  $-$0.6  &   neither  &   bright?  &  1 &  \ha-only        &    6.2  &  3.7 \\
21c$^f$  &  25.4  &  \phs\nodata  &  $-$0.9  &    \nodata  &   bright?  & 1 &  \ha-only  &   5.8  &  1.4 \\

43f  &  17.7  &  \phs 2.0  &  \phs 0.0  &     red  &    bright  &  1   &    FGND  &  0.011  &  \nodata \\
22a  &  19.6  &  \phs 2.5  &  $-$0.4  &       red  &    bright  &   0  &   FGND  &   0.021  &   \nodata  \\
51e$^f$  &  19.8  &  \phs 2.1  &  $-$0.1  &       red  &    bright  &   1  &   FGND  &   0.056 &  \nodata \\

52b$^f$  &  20.9  &  \phs 0.8  &  \phs 0.4  &      blue  &     faint  &   0  &    AGN  &  0.068 & \nodata \\
22f$^f$  &  21.6  &  \phs 2.3  &  \phs 0.2  &       red  &    faint?  &  1 &      AGN  &  0.57   &  \nodata \\
54d$^f$  &  22.2  &  \phs 0.7  &  \phs 0.3  &      blue  &     faint  &  0 &      AGN  &  8.8   &    \nodata \\
31b  &  22.9  &  \phs 1.4  &  \phs 0.2  &      blue  &   neither  & 1 &     AGN  &     0.74   &    \nodata \\
54g$^f$  &  23.7  &  \phs 2.6  &  \phs 0.0  &   neither  &   neither  & 1  &   AGN  &   0.96 &  \nodata \\
23c  &  23.8  &  \phs\nodata  &  \phs 0.4  &    \nodata  &    faint?  & 1 & AGN &     2.6 &    \nodata \\
62b  &  24.2  &  \phs 1.5  &  \phs 0.3  &      blue  &    faint?  &  0  &     AGN  &     18     & \nodata \\
41g  &  25.8  &  \phs 1.2  &  $-$0.9  &     blue?  &   bright?  &  1  &     AGN  &      39   &    \nodata \\

52c  &  20.7  &  \phs 1.1  &  \phs 0.2  &      blue  &     faint  &  0 &     AGN?  &  0.58 &  \nodata \\
33l  &  21.7  &  \phs 0.5  &  \phs 0.2  &      blue  &     faint  &   1 &    AGN?  &   3.4 &   \nodata \\
24g  &  22.2  &  \phs 0.6  &  \phs 0.3  &      blue  &     faint  &  1 &     AGN?  &   2.2 &  \nodata \\

33h  &  19.3  &  \phs 0.9  &  \phs 0.1  &      blue  &   neither  & 1 &  blue-only  &   0.20  &   \nodata \\    
44a  &  20.7  &  \phs 0.7  &  \phs 0.1  &      blue  &   neither  & 1 &  blue-only  &   2.2    &   \nodata \\   
51a  &  21.0  &  \phs 0.6  &  \phs 0.1  &      blue  &   neither  & 1 &  blue-only  &   0.78   &  \nodata \\     
33c  &  23.0  &  \phs 1.8  &  \phs 0.0  &     blue?  &   neither  & 1 &  blue-only  &  0.75  &   \nodata \\     
24a  &  23.8  &  \phs 2.1  &  \phs 0.0  &     blue?  &   neither  & 1 &  blue-only  &  1.5   &   \nodata \\    

\cline{1-10}%\\
\multicolumn{3}{l}{Alternate Counterparts:}\\
33e     &   23.4  &  1.8    &  $-$0.4 &  blue? &  bright  &   1   &  CV?              &   1.7 & 2.6 \\
					                  
21c    &    24.8 &  \ldots &  $-$0.6 &  \ldots &  bright? &   1   & Ha-only          & 3.3 &  1.4 \\
54g    &    25.2 &  \ldots &  $-$1.1 &  \ldots &  bright? &   1  &  Ha-only         &  3.8 & 1.2 \\
					                  
51e   &    15.9  &  1.7     &  \phs 0.1      &  red     &  bright? &   1  & FGND              & 0.002 & \nodata \\
54d   &    20.1 &  2.4     &  $-$0.2 &  red     &   bright  &   0  &  FGND             & 1.3 & \nodata \\
                 
52b   &    21.3 &  0.6     &  \phs 0.3       &  blue    &  faint    &   0  &  AGN               & 0.096 &  \nodata \\
44c   &    24.0 &  1.2     &  \phs 0.4       &  blue    &   faint   &   1  &  AGN?              & 3.5 &  \nodata \\
                  
21d   &   22.8  &  1.8     &  \phs 0.1       &  blue?   &  neither  &  1   &  blue-only       &  1.8 & \nodata \\
22f    &   23.6  &  1.7    &  $-$0.2  &  blue?   &  neither  &   1   & blue-only        & 3.6 & \nodata \\
23b   &   23.3  &  1.8    &  \phs 0.3        &  blue?   &  neither  &   1  &  blue-only       &  1.4 & \nodata \\

\enddata
\tablecomments{Photometric properties of the candidate optical counterparts, sorted by source type. Within each type, the sources are listed in decreasing \r-band brightness, except in the CV category where sources are listed as previously-known CVs first, then new CVs, then new CV?s.  The quality index is described in \S 2.3.
$^a$X-ray ID from HCD09.
$^b$Description of the candidate's location in the \r\ vs. \br\ CMD (see Fig. 3 and \S 3 for details).
$^c$Description of the candidate's location in the \r\ vs. \hr\ CMD (see Fig. 3 and \S 3 for details).
$^d$X-ray-to-optical flux ratio.  Unabsorbed X-ray fluxes from HCD09, revised downward by a factor of 1.25; \r-band fluxes assume \fr\ $= 10^{0.4R-5.89}$ (see \S 4.1).
$^e$X-ray luminosity calculated from the fluxes given by HCD09 assuming a distance of 4.9 kpc.
$^f$Alternate counterpart included in lower sub-table.
}
%\label{photom_tab}
\end{deluxetable}

\end{document}